\documentclass[sigconf,10pt]{acmart}

\usepackage{amsfonts}
\usepackage{algorithm}
\usepackage{algorithmic}
\usepackage[caption=false]{subfig}
\usepackage{url}
\usepackage{color}
\usepackage{xspace}
\usepackage{enumitem}
\usepackage{multirow}
\usepackage{makecell}
\usepackage{pifont}

\newcommand{\del}[1]{}
\newcommand{\add}[1]{#1}
\newcommand{\revise}[2]{#2}

\newcommand{\ours}{mzCache}
\newcommand{\oursit}{\emph{mzCache}}
\AtBeginDocument{%
  }

\copyrightyear{2026}
\acmYear{2026}
\setcopyright{cc}
\setcctype{by}
\acmConference[MobiCom '26]{The 32nd Annual International Conference on Mobile Computing and Networking}{October 26--30, 2026}{Austin, TX, USA}
\acmBooktitle{The 32nd Annual International Conference on Mobile Computing and Networking (MobiCom '26), October 26--30, 2026, Austin, TX, USA}
\acmDOI{10.1145/3795866.3844495}
\acmISBN{979-8-4007-2505-0/2026/10}

\author{Hongseung Yu$^{1}$, Minsung Kim$^{1}$, Jongseok Park$^{2}$, Kyunghan Lee$^{1}$}
\affiliation{%
  \institution{%
    $^{1}$Seoul National University \quad $^{2}$UC Berkeley\\
    \{hongseung, minsungkim.cs, kyunghanlee\}@snu.ac.kr, js\_park@berkeley.edu%
  }
  \country{}
}

\makeatletter
\let\originalspecialsection\@specialsection
\renewcommand{\@specialsection}[1]{%
  \originalspecialsection{\MakeUppercase{#1}}%
}
\makeatother

\renewcommand{\shortauthors}{Hongseung Yu, Minsung Kim, Jongseok Park, and Kyunghan Lee}
\renewcommand{\abstractname}{\MakeUppercase{Abstract}}

\begin{document}

\title{mzCache: On-Device LLM Memory Management under Multitasking}




\begin{abstract}

On-device mobile Large Language Model (LLM) inference is gaining significant attention. However, mobile devices operate in highly dynamic multitasking environments where users frequently switch between applications. This creates memory pressure, forcing LLM memory---model weights and KV cache---to be evicted by the operating system. When a new inference request arrives, the inference system must restore the evicted memory through slow storage reads or recompute the entire KV cache, severely degrading responsiveness. 
To address this, we present \textit{mzCache}, an on-device LLM inference system with specialized memory management for multitasking environments.
Under unpredictable memory pressure, mzCache elastically evicts LLM memory and leverages the unified memory of mobile SoCs to enable \textit{zero-wait} inference on the GPU with concurrent CPU-side restoration.
mzCache realizes this through restoration-oriented memory management: LLM memory is partitioned into fine-grained shared buffers to enable partial eviction and restoration with concurrent cross-processor access, while hybrid swap and backward-out eviction policies ensure low-latency restoration from any eviction state. Implemented on llama.cpp and deployed as an Android application, 
mzCache achieves 2.1--5.5$\times$ reduction in Time-to-First-Token compared to storage-backed partial offload and demonstrates its effectiveness in real multitasking scenarios.

\end{abstract}

\maketitle

\section{INTRODUCTION}

\begin{figure}[t] \centering \includegraphics[width=1\linewidth]{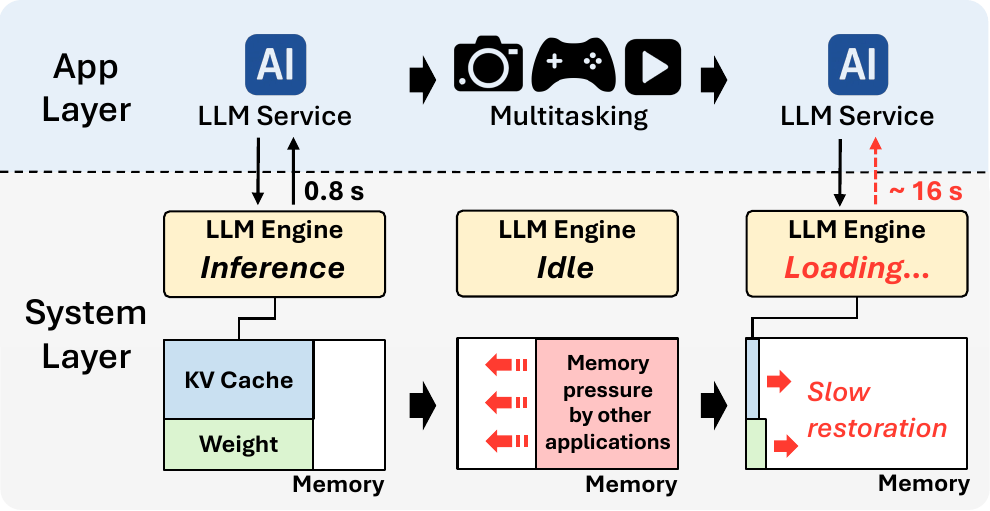} \caption{Multitasking on mobile devices: the user interacts with an on-device LLM service, switches to other apps, and later returns. Memory pressure from concurrent apps evicts LLM memory, increasing the response time from 0.8s to 16s.} \label{fig:problem} \end{figure}

Large Language Model (LLM)-based services are rapidly gaining adoption on mobile devices~\cite{Samsung2025-tp, Apple2025-eu, Huawei2025-ri, Android2025-vb}, unlocking personalized experiences that leverage local user data without cloud dependence.
On-device LLM assistants, for example, can summarize email threads using local message history or suggest contextually relevant responses based on prior conversations~\cite{li2025mobilora, zheng2024edge_llm, Chen2025-pg}.
These benefits stem from preserving the \textit{context} of prior interactions.

LLM inference systems achieve this by storing the interaction history as a set of key–value pairs known as the \textit{KV cache}.
Preserving the KV cache enables generating context-aware responses to new inputs while avoiding expensive recomputation.
Despite its benefits, as the KV cache accumulates over long or repeated interactions, even a single context for a small mobile-optimized model can grow to around 4 GB.
When combined with \textit{model weights}, the total LLM memory footprint reaches 5 to 7 GB, consuming nearly half of the typical 12 GB RAM available on modern smartphones~\cite{Samsung, iPhone17, Oneplus15}.

Such large memory demand poses a critical challenge in mobile \textit{multitasking environments}, where users frequently switch between multiple applications.
As illustrated in Figure~\ref{fig:problem}, after an LLM interaction completes, users transition to other applications such as camera, game, or video streaming.
As more apps are launched and used, the limited memory of mobile devices becomes rapidly scarce.
Under this memory pressure, the operating system (OS) evicts the idle LLM memory to make room for active applications.
When the user returns to the LLM service with a new inference request, this evicted memory must be restored before a response can be generated.
During this restoration, inference computation is blocked by slow storage reads and on-demand paging mechanisms~\cite{silberschatz2018operating}, substantially increasing Time-to-First-Token (TTFT)---the time to generate the first output token---from 0.8 to 16 seconds.

\begin{figure}[t] \centering \vspace{-3mm} \includegraphics[width=1\linewidth]{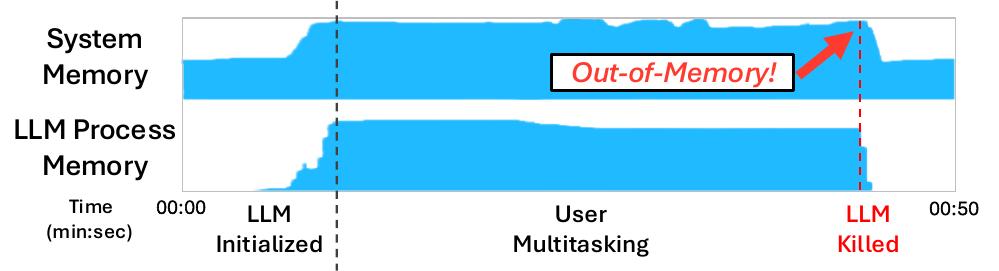} \caption{System and LLM process memory traces under multitasking, collected on a Galaxy S25+ using Perfetto~\cite{Android2025-xo}. As memory pressure increases, the Low-Memory Killer terminates the LLM process.} \label{fig:intro_trace} \end{figure}

To overcome the limited bandwidth of storage-only restoration, mobile OSes such as Android employ zRAM~\cite{Android2025-yo}, a compression-based in-memory swap space that compresses evicted pages within RAM itself.
However, zRAM's general-purpose compression algorithms (e.g., lz4~\cite{lz4Unknown-kw} and zstd~\cite{facebookUnknown-wy}) barely compress the KV cache, leaving memory pressure largely unrelieved.
As shown in Figure~\ref{fig:intro_trace}, this unresolved pressure leads to process termination by the Low-Memory Killer.
As a result, before generating a new response, the LLM system must reload weights from storage and recompute the entire KV cache.
Both operations incur significant latency, severely degrading the responsiveness and user experience.

As such, general OS memory management mechanisms are largely unsuitable
for LLMs. Unfortunately, existing mobile LLM services~\cite{yi2023mllm,
mlc-llm, Wang2024-or} focus primarily on runtime inference optimizations,
and their memory management is entirely dependent on these OS-based
mechanisms. As a result, the TTFT and user experience of LLM-based
applications suffer greatly in multitasking mobile environments, calling
for an LLM- and mobile-aware system that properly handles external memory
pressure.

To this end, we present \oursit{}, a complete on-device LLM inference system that
integrates specialized memory management for responsive inference under multitasking. A key challenge is that memory pressure
from competing applications is inherently dynamic and unpredictable. To
handle this, \ours{} employs a restoration-oriented design, elastically
reducing its memory footprint according to external pressure while
maintaining readiness for rapid restoration throughout the eviction
process.

\ours{} first establishes a memory management mechanism tailored to mobile platforms: it partitions weights and KV cache into fine-grained units and allocates them as shared buffers on the mobile SoC's unified memory. This enables evicting only the necessary amount of memory while allowing the CPU to perform restoration and the GPU to execute inference without redundant data transfers between processors. On top of this, \ours{} introduces two policies that govern where and when each unit should be evicted, so that any partial eviction state allows for the fastest possible restoration regardless of the eviction amount.

To overcome the storage read bottleneck in restoration, we draw
inspiration from Android's zRAM and apply KV cache compression
techniques~\cite{Liu2024-wu, Hooper2024-yl, Liu2024-kivi, Sheng2023-cy,
Zhang2024-cq} that exploit its structural properties, making in-memory
compressed space a viable second restore path alongside storage reads.
Evicted KV cache is distributed across the two paths so that both remain
fully utilized during restoration. Beyond placement, \ours{} evicts later
layers first so that early layers, which LLM inference executes first, are
always preserved. During restoration, the GPU can begin inference
immediately on the preserved front layers while the CPU restores evicted
layers in the same forward order, maximizing the overlap between
restoration and inference.

We implement \ours{} on top of llama.cpp~\cite{Cpp2023-tx}, a widely used LLM inference framework for mobile environments, and evaluate its performance across various mobile devices and LLM models.
Compared to a storage-backed partial offload baseline, \ours{} reduces TTFT by 2.1--5.5$\times$ and demonstrates effectiveness in real-world multitasking scenarios under memory pressure.
The main contributions of this paper are summarized as follows:
\begin{itemize}[topsep=0pt, partopsep=0pt, itemsep=0pt, parsep=0pt, leftmargin=2em]
\item We reveal that OS memory management is inadequate for LLM workloads in mobile multitasking environments, causing severe TTFT degradation.
\item We present \ours{}, a complete LLM inference system with restoration-oriented memory management that enables responsive inference from any eviction state.
\item We test \ours{} on commercial smartphones, show 2.1–5.5× faster TTFT than partial offload, and validate its practicality under realistic multitasking pressure.
\end{itemize}
\section{BACKGROUND}
\subsection{Characteristics of LLM}
% Modern LLMs are constructed by stacking dozens of transformer~\cite{vaswani2017attention, han2021transformer} layers, where the output of each layer is sequentially fed into the next.
Modern LLMs are constructed by stacking dozens of transformer layers~\cite{vaswani2017attention}.
During inference, input tokens are processed through these stacked layers to generate output tokens.
The inference consists of two stages: First, the \textit{prefill} stage processes the entire input sequence of tokens.
For each token, every transformer layer computes key and value (KV) vectors, and the resulting KV pairs are stored as \textit{KV cache}.
Because self-attention in this stage requires interactions across all input tokens, the prefill latency, referred to as Time-to-First-Token (TTFT), grows quadratically with the input sequence length.
The second stage, called the \textit{decode} stage, autoregressively generates output tokens one by one.
In the decode stage, each newly generated token references the accumulated KV cache to avoid recomputation, making each decoding step far cheaper than prefill.

\subsection{Context-Aware LLM Inference in Mobile}
In mobile LLM inference, preserving the KV cache across interactions is particularly important, as applications are designed to deliver personalized experiences.
For instance, mobile LLM services process private user data—such as message history, email threads, and chat logs—as inference context, storing it as KV cache~\cite{li2025mobilora, zheng2024edge_llm, Chen2025-pg}.
The model then references this KV cache during inference to generate relevant responses across  interactions.

\begin{table}[]
\caption{Memory footprint of representative LLMs used in mobile and edge environments. 
All models listed here have fewer than 3 billion parameters.}
\centering
\large
\resizebox{\columnwidth}{!}{
\begin{tabular}{ccccc} \toprule
\textbf{LLM model} &
  \textbf{Weight} &
  \textbf{\begin{tabular}[c]{@{}c@{}}Context Length\\ (max)\end{tabular}} &
  \textbf{\begin{tabular}[c]{@{}c@{}}KV\\ (max)\end{tabular}} &
  \textbf{\begin{tabular}[c]{@{}c@{}}Max footprint\\ (Weight + KV)\end{tabular}} \\ \hline\hline
EXAONE-4.0-1.2B~\cite{exaone-4.0}   & 2.6 GB & 64k   & 3.8 GB  & \textbf{6.4 GB}  \\ \hline
Gemma2-2B~\cite{Gemma-Team2024-lv} & 4.9 GB & 8k   &  0.8 GB & \textbf{5.7 GB}  \\ \hline
Llama3.2-1B~\cite{dubey2024llama}    & 2.5 GB & 128k & 4 GB  &  \textbf{6.5 GB}   \\ \hline
Qwen3-0.6B~\cite{Qwen-Team2025-ry} & 1.2 GB & 32k   & 3.5 GB & \textbf{4.7 GB} \\ \hline
Qwen3-1.7B~\cite{Qwen-Team2025-ry} & 3.5 GB & 32k   & 3.5 GB & \textbf{7.0 GB} \\ \bottomrule
\end{tabular}
}
\label{table:context_size}
\end{table}

Preserving the KV cache, however, directly affects the memory footprint. As shown in Table~\ref{table:context_size}, for models under 3 billion parameters, the KV cache alone can grow up to 4 GB depending on context length, reaching 5--7 GB when combined with model weights.
In practice, reaching such long contexts is common: popular LLM datasets show that a single interaction produces hundreds of tokens on average~\cite{ShareGPT2022-wf, Yan2025-sc, Zheng2023-cn}, and contextual input such as email threads easily exceeds a thousand tokens~\cite{Xu2025-ji}. 
As these interactions accumulate throughout the day, the context quickly approaches its maximum.

The large KV cache also affects the computational cost of prefill.
When a new request arrives, the model must process all new input tokens while performing attention over the entire stored KV cache.
This causes the prefill computation to scale with both the new input length and the size of the accumulated KV cache, thus increasing the TTFT.

\add{
\subsection{Memory Pressure under Multitasking}
\label{section:memory_pressure_under_multitasking}
Modern mobile devices operate in highly dynamic multitasking environments, where users switch between applications over 100 times per day~\cite{Deng2019-oz}. As multiple apps run concurrently, they create unpredictable memory pressure on the limited RAM available on mobile devices. Under such pressure, the mobile operating system (OS) must reclaim memory from inactive processes, such as background applications that the user is not actively using, to make room for active applications~\cite{Android2025-yo}.

To reclaim memory under pressure, the OS relies on two
mechanisms. The first is \textit{paging}, which evicts pages from RAM: file-backed pages are simply dropped because their contents already reside on storage, while anonymous pages have no such backing and must be swapped out to a swap space. If the swap space resides in flash storage as a swap file, the high latency of storage reads during restoration severely degrades responsiveness. To address this, modern
mobile OSes such as Android employ zRAM~\cite{Android2025-yo}, an
in-memory compressed swap space that stores evicted pages in
compressed form within RAM itself. This trades CPU cycles for faster restoration than storage-backed swap. When memory pressure persists despite paging, the OS escalates to its second mechanism: the \textit{Low-Memory Killer (LMK)}, which terminates background processes to free their entire memory.
}
\section{\MakeUppercase{Motivation}}
\subsection{\revise{Memory Management in Mobile OS}{Limitations of OS Memory Management for LLMs}}
\label{section:memory_management_in_mobile_os}

\begin{table}[]
\caption{KV cache space savings under lz4 and zstd at an 8k-token WikiText context. These zRAM algorithms achieve only 0.2--9.3\% space savings, indicating KV cache compresses poorly under zRAM.}
\label{table:compress}
\resizebox{\columnwidth}{!}{
\begin{tabular}{cccccc}
\toprule
 & \textbf{\begin{tabular}[c]{@{}c@{}}EXAONE-4.0\\ 1.2B~\cite{exaone-4.0}\end{tabular}} & \textbf{\begin{tabular}[c]{@{}c@{}}Gemma2\\ 2B~\cite{Gemma-Team2024-lv}\end{tabular}} & \textbf{\begin{tabular}[c]{@{}c@{}}Llama3.2\\ 1B~\cite{dubey2024llama}\end{tabular}} & \textbf{\begin{tabular}[c]{@{}c@{}}Qwen3\\ 0.6B~\cite{Qwen-Team2025-ry}\end{tabular}} & \textbf{\begin{tabular}[c]{@{}c@{}}Qwen3\\ 1.7B~\cite{Qwen-Team2025-ry}\end{tabular}} \\ \hline\hline
\textbf{lz4 (\%)} & 0.28 & 0.20 & 1.19 & 0.19 & 0.19 \\ \hline
\textbf{zstd (\%)} & 9.30 & 8.65 & 9.28 & 8.04 & 8.04 \\ \bottomrule
\end{tabular}
}
\end{table}

\add{Depending on how the inference engine allocates memory, LLM
data resides in one of two memory regions---\emph{pageable} or
\emph{page-locked}---and under multitasking memory pressure, the OS
handles neither region well.}

\begin{figure}[t]
    \centering
    % \vspace{-3mm}
    \includegraphics[width=\columnwidth]{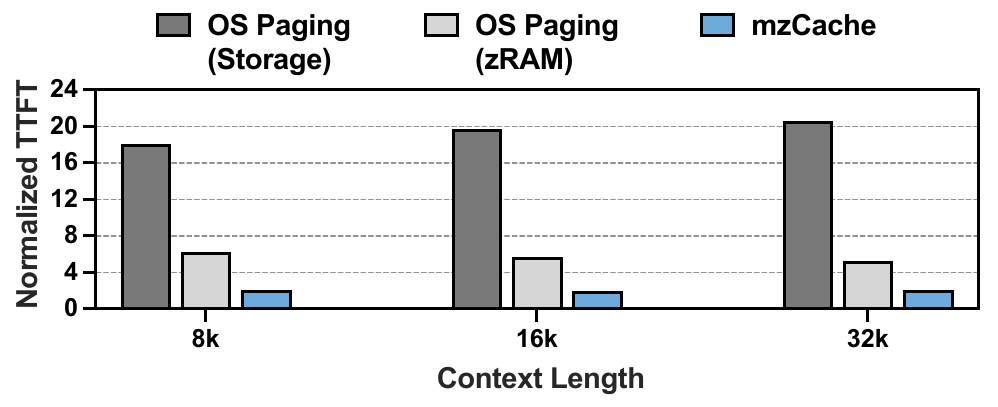}
    \caption{TTFT under full eviction (Galaxy S25+, Qwen3-0.6B, 8 input tokens), normalized to the all-in-memory case. mzCache keeps TTFT close to the in-memory case. Note that zRAM still leaves memory pressure largely unrelieved (Table~\ref{table:compress}).
    }
    \label{fig:swap_slow}
\end{figure}

\add{\emph{1) Pageable region.} The OS can reclaim this region
through paging, but with mechanisms blind to LLM data.}
For LLM processes, file-backed pages containing model weights are dropped, while anonymous pages containing the KV cache are written to zRAM.
However, zRAM uses general-purpose compression algorithms such as lz4~\cite{lz4Unknown-kw} or zstd~\cite{facebookUnknown-wy}, which rely on byte-level pattern matching. These algorithms fail on KV cache because it consists of arithmetic data that lacks such patterns: as shown in Table~\ref{table:compress}, space savings range from only 0.2\% to 9.3\% across different LLM models.
As a result, memory pressure remains largely unrelieved, as the poorly compressed KV cache continues to occupy substantial memory in zRAM.
\revise{When memory pressure persists despite swapping, the
mobile OS takes another approach: the Low-Memory Killer (LMK),
which terminates background processes to free memory. Due to
their large memory consumption, LLM processes are particularly
vulnerable to such termination~\cite{Android2023-ma}. When the
LMK terminates the LLM process,}{When such pressure persists,
the LMK eventually terminates the LLM process, whose large
memory footprint makes it particularly
vulnerable~\cite{Android2023-ma}. Upon termination,} the
entire KV cache representing the user's context is lost.
Resuming the service requires a cold-start that reloads
weights from storage and recomputes the KV cache from
scratch, leading to severe responsiveness degradation.

Even when the LLM process avoids termination, the
restoration introduces substantial latency. When an inference request arrives, the OS must restore both weights from storage and KV cache from swap space through on-demand paging: pages are brought back into memory only when accessed during inference, triggering page faults~\cite{silberschatz2018operating}. 
Because prefill must access all weights and the entire KV cache to generate the first token, page faults continue throughout restoration until all evicted data has been recovered. 
As shown in Figure~\ref{fig:swap_slow}, this on-demand restoration severely degrades TTFT.
Under full eviction, TTFT increases by 18--20 times when using flash storage swap and 5--6 times when using zRAM, compared to the all-in-memory case.

\add{\emph{2) Page-locked region.} Memory pinned by an
accelerator (GPU or NPU) driver, or locked through system calls like \texttt{mlock}, is exempt from
paging: the OS cannot evict these pages regardless of
pressure. The only lever the OS retains here is the LMK, which terminates the process and thereby incurs the cold-start penalty described above.}

To avoid even this penalty, Android
AICore~\cite{Android2025-vb} provides Google's Gemini Nano
model as a system-level service effectively shielded from LMK
termination~\cite{Android2025-yo}, with its weights locked in
memory and exempt from paging~\cite{Triggs2025-le,
Triggs2024-cf}. As a result, the weights always remain in
memory, ready for inference without restoration. This
protection, however, comes at a critical cost: the locked
weights occupy RAM even when the service is idle, wasting
scarce memory and reducing headroom for other applications.
Moreover, it provides no solution for the KV cache, which can
still be evicted under pressure, reintroducing significant
latency upon resumption.

\del{Given that mobile OS mechanisms fail to manage LLM memory
effectively, mobile LLM inference systems must incorporate
principled memory management to handle memory pressure in
multitasking environments.}

\begin{figure}[t]
    \centering
    % \vspace{-3mm}
    \includegraphics[width=\linewidth]{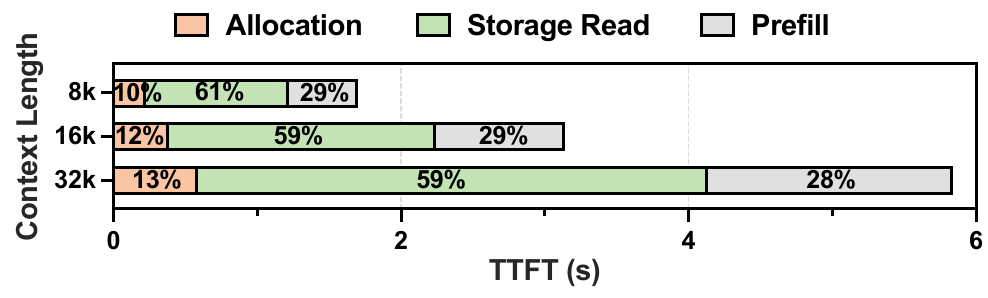}
    \caption{TTFT breakdown when loading entire weights and KV cache from flash storage and executing inference on GPU (OnePlus 12, Qwen3-0.6B, 8 input tokens), showing time spent on allocation, storage reads, and prefill across different context lengths.}
    \label{fig:alloc_read_prefill}
\end{figure}

\subsection{Design Constraints}
\label{section:platform_design_constraints}

\add{Since the OS handles neither memory region well, \emph{mobile LLM inference systems must themselves incorporate principled memory management.}}
\revise{Building such a memory management solution}{Building
such a solution} requires two capabilities: (1) evicting memory under pressure to make room for concurrent applications, and (2) rapidly restoring evicted data to run inference when a new request arrives. However,
several constraints in existing LLM inference systems and mobile
platforms make achieving these capabilities challenging.

\noindent\textbf{Monolithic memory buffer.} LLM inference systems typically adopt monolithic memory allocation: they allocate one large contiguous buffer at initialization to avoid repeated allocation overhead and simplify memory layout management for performance-critical kernels. However, the monolithic approach operates as all-or-nothing: the system cannot adjust its memory footprint in response to memory pressure. Without mechanisms to partially release or allocate memory, adaptive memory management becomes infeasible.

\noindent\textbf{Limited memory hierarchy.} On mobile platforms, prefill relies on accelerators such as GPUs to achieve acceptable performance~\cite{Xu2025-ji, Chen2025-fe}. When data is not resident in accelerator memory, the system must restore it before prefill can proceed. On servers, discrete GPU systems provide a separate host memory tier: data offloaded from GPU memory can be temporarily stored in host memory and quickly restored via high-bandwidth interconnects such as PCIe~\cite{Jeong2025-ll, Chen2025-ub, Gao2025-ey, Yu2025-kg}. Mobile platforms, however, employ unified memory architectures where accelerators and the CPU share the same physical memory, eliminating this intermediate fast-access tier. Consequently, systems must rely on flash storage, where slow read performance becomes a critical bottleneck. As shown in Figure~\ref{fig:alloc_read_prefill}, storage reads dominate restoration time, accounting for 59--61\% of TTFT across different context lengths.

\noindent\textbf{Allocation overhead.} Since eviction releases memory regions back to the system, restoration requires re-allocating them before any data can be loaded. This re-allocation incurs substantial overhead, as it involves system calls that trigger page allocation, zeroing, and IOMMU / SMMU\footnote{IOMMU (I/O Memory Management Unit) and SMMU (System MMU) translate device virtual addresses to physical addresses for accelerators such as GPUs and NPUs.} page table updates~\cite{ARM2025-hw, Hawkes2020-zf}. The cost scales with the allocation size and recurs at every restoration. As shown in Figure~\ref{fig:alloc_read_prefill}, allocation accounts for 10--13\% of TTFT, representing a non-trivial cost that compounds the restoration latency.

\subsection{Opportunities}
\label{section:opportunities}

To achieve the goal of responsive LLM inference restoration, it is critical to understand and leverage the unique characteristics of mobile memory systems. We identify two such characteristics: (1) the use of compressed memory regions, and (2) the unified memory architecture.

Due to the low access speeds of flash storage on mobile devices, memory offloading on mobile is often bound by storage I/O. To overcome this, Android's zRAM offers a compressed region in the main memory, providing a small but high-throughput alternative to the larger but slower storage path (§\ref{section:memory_pressure_under_multitasking}). We reinterpret this dual-choice offloading concept of the mobile OS for LLM memory management, and apply compressed offloading to the dynamically generated KV cache while keeping the static model weights on the flash storage. The reason is clear: the KV cache exhibits similarity between adjacent tokens and channels, allowing higher compression ratios with minimal accuracy loss~\cite{Liu2024-wu, Hooper2024-yl, Liu2024-kivi, Sheng2023-cy, Zhang2024-cq}, while the larger model weights can remain unchanged in storage, allowing reduced I/O on eviction and a more efficient use of the system's memory resources.  

The single shared physical memory on mobile devices also provides a key opportunity for efficient LLM memory management. This unified memory provides a common workspace for different processors of the mobile SoC to coordinate the use and management of the LLM memory, often at the same time. For example, the GPU can focus on high-throughput prefill computation, while the CPU handles memory management tasks such as allocation, storage reads, and decompression, without requiring costly memory copies between the two processors. By carefully partitioning the LLM memory into manageable chunks and scheduling the access patterns between the GPU and CPU, the two processors can cooperate to leverage the unified memory of the mobile system to its full potential.

\add{A natural concern in exploiting both opportunities simultaneously is memory bandwidth contention, as the two restore paths and the two processors share the physical memory subsystem. In practice, however, this shared bandwidth is an underutilized resource, since no single processor can fully utilize it~\cite{Chen2025-fe} and storage reads occupy only a small fraction of it\footnote{\add{On Galaxy S25+, the sequential-read bandwidth of UFS~4.0 (4.2\,GB/s) is far below the peak bandwidth of LPDDR5X (84.8\,GB/s).}}. The shared memory bandwidth is thus an enabler rather than a bottleneck: engaging the CPU, GPU, and storage together puts this spare bandwidth to work for restoration.}

\begin{figure}[t]
    \centering
    \includegraphics[width=1\linewidth]{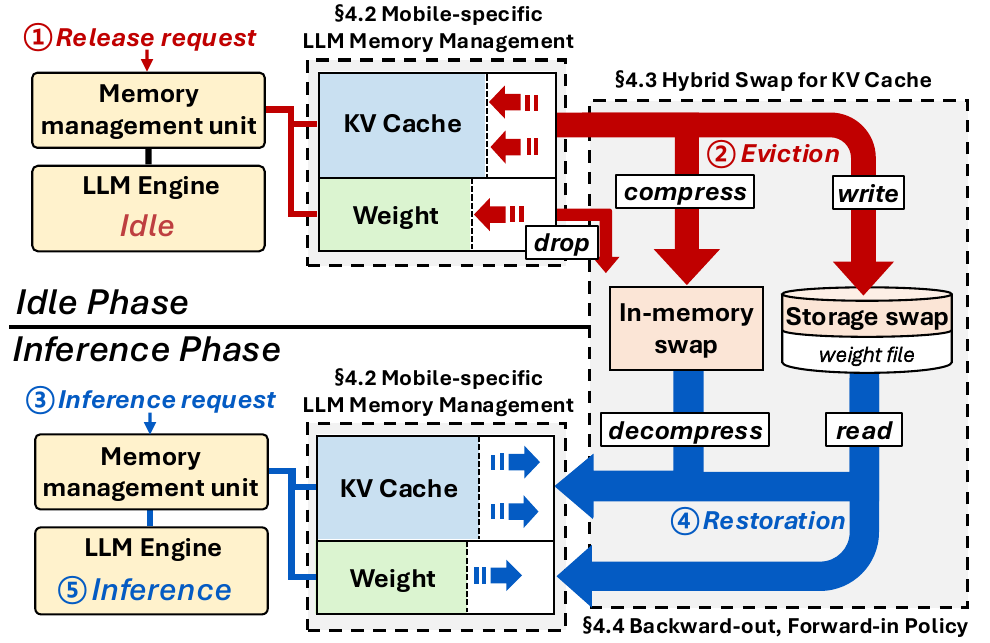}
    \caption{System Overview of \ours{}}
    \label{fig:system_overview}
\end{figure}
% \vspace{-0.5em}

\section{\MakeUppercase{System Design}}

\subsection{Overview}
\label{section:overview}

\newcommand{\circled}[1]{\ding{\numexpr191+#1\relax}}
Figure~\ref{fig:system_overview} presents an overview of \oursit{}. \ours{} operates in two distinct phases: the \textit{idle phase} and the \textit{inference phase}. During the idle phase, when the system receives a memory release request due to external memory pressure~\circled{1}, \ours{} evicts the requested amount of LLM memory~\circled{2}. The idle phase continues until a new inference request arrives~\circled{3}. During the inference phase, \ours{} restores evicted LLM memory~\circled{4} and performs inference~\circled{5}. After the inference completes, the system transitions back to the idle phase.

Unlike server-side LLM systems with exclusive control over GPU memory, a mobile LLM inference system faces \emph{dynamic and unpredictable} pressure from competing user applications. To overcome this novel challenge, it is critical that LLM memory is elastically reduced according to outside memory pressure, evicting only the necessary amount of memory
while maintaining readiness for rapid restoration throughout the eviction process. \ours{} enables and leverages this \textit{\textbf{Restoration-oriented Eviction}} concept to the fullest, allowing low-latency restoration of LLM memory under any pattern or degree of memory pressure that an LLM inference system may face in the mobile environment.

To realize this, we first design an LLM memory management mechanism that can leverage the unified memory of mobile environments to enable fine-grained management of LLM memory with concurrent model execution (§\ref{section:finegrained}). On top of this mechanism, two policies govern the eviction and restoration process to ensure efficient memory recovery from any intermediate eviction state, providing a robust solution to the problem of dynamic memory pressure in the mobile environment.
The first policy determines \textit{where} to place evicted data across the two restore paths, in-memory decompression and storage reads, to balance restoration load between the two paths (§\ref{section:dual_path}). The second policy determines \textit{when} each LLM data should be evicted and restored, to minimize the time to first token generation in LLM inference restoration (§\ref{section:restore_inference_overlap}).

% \vspace{-0.3em}
\subsection{Mobile-specific LLM Memory Management}
\label{section:finegrained}

The main goal of \ours{}'s memory management mechanism is to bridge the gap between mobile memory systems and its restoration-oriented eviction policies. To achieve this, we first depart from the monolithic allocation of traditional LLM inference systems and partition the layer weights and KV cache into finer-grained units, allowing the system to elastically evict data only to the minimum required amount.
We use layer-level granularity for weights, since each layer forms a fixed-size unit that serves as a natural boundary for model execution. For KV cache, the layer-level granularity is insufficient as the KV cache size varies with context length. We instead divide each layer's KV cache into fixed-size chunks following popular server-side practices~\cite{Kwon2023-ci}, enabling flexible partial eviction regardless of context length. We detail the exact chunk size of the KV cache in §\ref{section:dual_path}, as the most effective size depends on the I/O characteristics of the given memory system.

\begin{figure}[t]
\centering
\includegraphics[width=\linewidth]{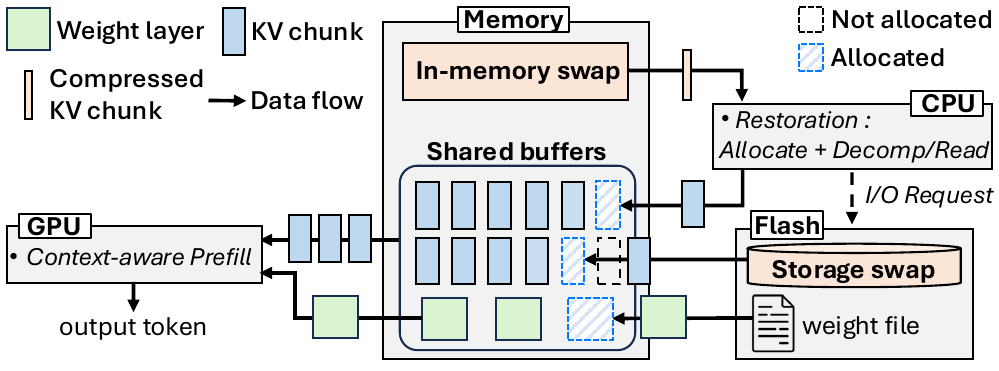}
\caption{Overview of GPU-CPU cooperation on the mobile SoC's unified memory during \ours{} restoration.}
\label{fig:cpu_gpu_mem_flash}
\end{figure}

Using Figure~\ref{fig:cpu_gpu_mem_flash} as an example, we show how \ours{} leverages the unified memory and fine-grained units to achieve low-latency LLM memory restoration on mobile devices. We first allocate the weight layers and KV cache chunks (KV chunks) as \textit{shared buffers} directly visible to both the CPU and the GPU. When a new inference request arrives after a partial eviction, \ours{} immediately begins the prefill computation of the request on the GPU using the partial weights and KV chunks that remain in shared buffers. Concurrently, \ours{} also restores partially evicted data on the CPU by loading weight layers and KV chunks from storage via direct I/O and decompressing KV chunks from in-memory swap, all directly into shared buffers. Since each fine-grained memory unit can be independently allocated and loaded, we distribute memory allocation and data loading across multiple CPU cores to maximize restoration throughput. This design removes any explicit data transfers between the two processors and allows seamless pipelining between the restoration on the CPU and the consumption of the restored data on the GPU to achieve minimum-latency LLM memory restoration in mobile environments.

To compute directly on KV chunks residing in noncontiguous shared buffers, \ours{} employs a custom attention kernel using OpenCL~\cite{KhronosGroup-opencl}. Unlike server-side systems that must coordinate KV cache history across concurrent multi-user requests using a prefix-tree system~\cite{Kwon2023-ci, Yu2025-kg, Zheng2024-sglang}, mobile inference serving a single user can manage KV cache through a simple array of KV chunk pointers. Our kernel targets prefill over a large KV cache accumulated from prior interactions. Due to the restricted load/store bandwidth of mobile GPU on-chip memory~\cite{Lam2024-adreno} and limited data reuse from single-batch serving of mobile LLMs, we find that caching the KV data on the GPU on-chip memory is often slower than streaming directly from shared buffers.
Therefore, the kernel reads KV chunks directly from shared buffers and streams through the chunk pointer array, applying online softmax~\cite{Dao2022-pj} across noncontiguous chunks to enable seamless single-pass attention without consolidation. To maximize throughput, the kernel employs 8-wide FP16 vector operations that exploit the double-rate FP16 execution hardware of the Qualcomm Adreno GPUs that we mainly target.

\subsection{Hybrid Swap for KV Cache}
\label{section:dual_path}

As identified in §\ref{section:opportunities}, compressing the KV cache in memory during eviction opens a second restore path, in-memory decompression, alongside storage reads. However, a fixed assignment that simply directs the KV cache to in-memory space and the weights to storage is insufficient: the two restore paths have different throughputs and the KV cache size varies with context length, so restoration inevitably becomes imbalanced. To address this, \ours{} introduces \textit{hybrid swap} for the KV cache: rather than assigning KV cache exclusively to in-memory compressed space, a portion is also written to storage, enabling the system to distribute data across both paths.

The key question is how to determine this distribution, given that the system cannot know in advance how much memory will ultimately be evicted.
Consider, for example, a policy that evicts all KV chunks first across both paths before dropping any weight layer. Distributing KV cache evenly may appear balanced initially. But once all KV cache has been evicted and further eviction is needed, the system has no choice but to drop weight layers, which must later be reloaded from storage. As more weight layers are dropped, the restoration load shifts increasingly toward the storage path, breaking the balance. This illustrates that a more principled approach is needed.

\begin{figure}[t]
    \centering
    \includegraphics[width=\linewidth]{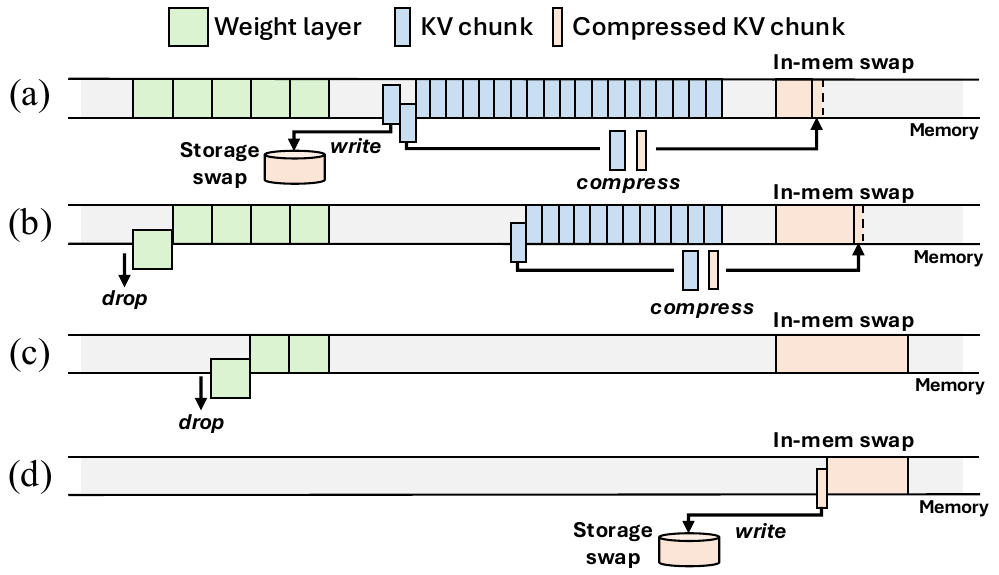}
    \caption{Four eviction stages in mzCache. (a) \textsc{KVonly}: evict KV chunks to both swap spaces. (b) \textsc{KVandW}: drop weights while evicting KV. (c) \textsc{Wonly}: drop only weights. (d) \textsc{CompKV}: compressed KV may be evicted to storage under further pressure.}
    \label{fig:four_stages}
\end{figure}

\noindent\textbf{Offline profiling.} Balancing the two restore paths requires knowing how long each takes to process a unit of data. \revise{\ours{} profiles three per-unit times on each target device: $t_{d}^{\mathrm{KV}}$, the time to decompress one KV chunk; $t_{r}^{\mathrm{KV}}$, the time to read one KV chunk from storage; and $t_{r}^{\mathrm{W}}$, the time to read one weight layer from storage. This profiling assumes that total restoration time scales linearly with the number of units.}{These per-unit times come from a one-time profiling on each target device, which measures two throughputs: the decompression throughput of the configured compression algorithm and the sequential read throughput of the flash storage via direct I/O. Given the sizes of a KV chunk and a weight layer, dividing each by the corresponding throughput yields the three per-unit times: $t_{d}^{\mathrm{KV}}$ (decompress one KV chunk), $t_{r}^{\mathrm{KV}}$ (read one KV chunk from storage), and $t_{r}^{\mathrm{W}}$ (read one weight layer from storage). Even when a different model changes these unit sizes, the per-unit times are simply re-derived from the same measured throughputs, without re-profiling the device.}

\noindent\add{\textbf{KV chunk size.} For the per-unit profiling above to be valid, each path must operate at full throughput even for a single unit. Decompression satisfies this regardless of data size: in our measurements, its throughput remains constant.} \revise{For storage reads, each KV chunk must be large enough to saturate the storage bandwidth in a single read, so that the time to read $N$ chunks scales linearly with $N$.}{Storage reads, in contrast, reach full bandwidth only when each read request is sufficiently large:} in our experiments on UFS 4.0 flash storage, sequential reads saturate the storage bandwidth at about 512 KB per operation.\add{ Weight layers, at tens of megabytes each, are well above this size.} To meet this threshold for models with a small KV hidden dimension (e.g., 512), a minimum of 256 tokens per chunk is required.\footnote{$256 \times 512 \times 2\text{ (K, V)} \times 2\text{ bytes (FP16)} = 512$ KB.} We therefore fix the chunk size at 256 tokens across all configurations.

\noindent\textbf{Four-stage eviction.} Since KV chunks can be directed to either restore path, they serve as the balancing lever between the two paths. \revise{The key quantity is the threshold $T$: the number of KV chunks the decompression path needs to stay occupied for the full duration of weight restoration from storage.}{To keep the two paths balanced, \ours{} sets aside enough KV chunks to cover the weight reload: the threshold $T$ is the number of KV chunks that keeps the decompression path occupied while all weight layers are reloaded from storage.} The eviction policy follows from how the remaining number of KV chunks ($N_{\mathrm{KV}}$) relates to $T$ \revise{as eviction progresses}{throughout eviction}.

\renewcommand{\algorithmicrequire}{\textbf{Input:}}
\renewcommand{\algorithmiccomment}[1]{\hfill$\triangleright$~#1}
\begin{algorithm}[t]
\caption{Hybrid Swap Eviction Policy}
\label{alg:eviction}
\begin{algorithmic}[1]
\REQUIRE Per-unit times $t_{d}^{\mathrm{KV}}$, $t_{r}^{\mathrm{KV}}$,
         $t_{r}^{\mathrm{W}}$; threshold $T$

\STATE $N_{\mathrm{KV}} \gets N_{\mathrm{KV}}^{\mathrm{total}}$
       \COMMENT{KV chunks not yet evicted}
\STATE $N_{\mathrm{W}} \gets N_{\mathrm{W}}^{\mathrm{total}}$
       \COMMENT{weight layers not yet dropped}

\WHILE{release request}
    \IF{$N_{\mathrm{KV}} > T$}
        \STATE $N_{\mathrm{KV}} \gets$
        \textsc{KVonly}
            ($N_{\mathrm{KV}},
             t_{d}^{\mathrm{KV}},
             t_{r}^{\mathrm{KV}}$)
    \ELSIF{$N_{\mathrm{KV}} > 0$ \AND $N_{\mathrm{W}} > 0$}
        \STATE $N_{\mathrm{KV}}, N_{\mathrm{W}} \gets$
        \textsc{KVandW}
            ($N_{\mathrm{KV}},
             N_{\mathrm{W}},
             t_{d}^{\mathrm{KV}},
             t_{r}^{\mathrm{W}}$)
    \ELSIF{$N_{\mathrm{W}} > 0$}
        \STATE $N_{\mathrm{W}} \gets$
        \textsc{Wonly}($N_{\mathrm{W}}$)
    \ELSE
        \STATE \textsc{CompKV}()
    \ENDIF
\ENDWHILE
\end{algorithmic}
\end{algorithm}

Let $N_{\mathrm{KV}}^{\mathrm{total}}$ and $N_{\mathrm{W}}^{\mathrm{total}}$ denote the total number of KV chunks and \add{evictable }weight layers, respectively.\footnote{\add{We set $N_{\mathrm{W}}^{\mathrm{total}}$ to the total number of weight layers, allowing full eviction; setting it to 0 would keep all weights resident in memory.}} \revise{From the profiled values, we derive:}{While reloading one weight layer from storage, the decompression path can process $\lfloor t_{r}^{\mathrm{W}} / t_{d}^{\mathrm{KV}} \rfloor$ KV chunks, giving:}
$$\add{T = N_{\mathrm{W}}^{\mathrm{total}} \cdot \lfloor t_{r}^{\mathrm{W}} / t_{d}^{\mathrm{KV}} \rfloor}$$
Two cases arise based on whether $N_{\mathrm{KV}}^{\mathrm{total}}$ exceeds $T$ (Algorithm~\ref{alg:eviction}, Figure~\ref{fig:four_stages}).

\del{When $N_{\mathrm{KV}}^{\mathrm{total}} > T$, the KV cache exceeds what the decompression path alone can handle. The system starts in \textsc{KVonly}, where KV chunks are distributed across both in-memory compressed space and storage at a ratio determined by $t_{d}^{\mathrm{KV}}$ and $t_{r}^{\mathrm{KV}}$, while weights remain untouched. Once $N_{\mathrm{KV}}$ falls to $T$ or below, the system transitions to \textsc{KVandW}, compressing remaining KV chunks to in-memory space while dropping weight layers, with the pace guided by $t_{d}^{\mathrm{KV}}$ and $t_{r}^{\mathrm{W}}$. Because $N_{\mathrm{KV}}$ equals $T$ at this transition, KV chunks and weight layers are exhausted at a similar pace without entering \textsc{Wonly}.}

\add{\emph{1) \textsc{KVonly} $\rightarrow$ \textsc{KVandW} ($N_{\mathrm{KV}}^{\mathrm{total}} > T$).} The KV cache exceeds what the decompression path alone can handle, so eviction starts in \textsc{KVonly}, evicting KV chunks to both swap spaces while weights remain untouched. The chunks are split in inverse proportion to $t_{d}^{\mathrm{KV}}$ and $t_{r}^{\mathrm{KV}}$, with the faster path taking more, so that both paths finish restoring their shares in the same amount of time. Once $N_{\mathrm{KV}}$ falls to $T$, the system transitions to \textsc{KVandW}, compressing remaining KV chunks to in-memory space while dropping weight layers: for every layer dropped, $\lfloor t_{r}^{\mathrm{W}} / t_{d}^{\mathrm{KV}} \rfloor$ KV chunks are compressed. Because $N_{\mathrm{KV}}$ equals $T$ at this transition, the remaining KV chunks and the weight layers run out together.}

\del{When $N_{\mathrm{KV}}^{\mathrm{total}} \leq T$, all KV cache fits within the decompression path. The system starts directly at \textsc{KVandW}. After all KV chunks are exhausted, remaining weight layers are dropped in \textsc{Wonly}.}

\add{\emph{2) \textsc{KVandW} $\rightarrow$ \textsc{Wonly} ($N_{\mathrm{KV}}^{\mathrm{total}} \leq T$).} All KV cache fits within the decompression path, so eviction starts directly at \textsc{KVandW}. With fewer KV chunks than the weight reload requires, however, the KV cache is exhausted while weight layers still remain. Eviction then continues in \textsc{Wonly}, dropping the remaining weight layers.}

\del{In both cases, if further pressure arrives after all data has been evicted, the system enters \textsc{CompKV}, which moves compressed KV from in-memory space to storage to release additional memory.}

\add{In either case, once all weight layers have been dropped and all KV chunks evicted, only the compressed KV chunks remain in the in-memory swap space. If memory pressure persists even then, the system enters \textsc{CompKV}, moving the compressed KV to storage to release this last portion of memory.}

\del{\textsc{Wonly} and \textsc{CompKV} extend beyond the balanced region, but because they are entered only after the earlier stages have already distributed KV chunks across both paths, the system preserves the best achievable balance for the given eviction amount.}

\noindent\add{\textbf{Balance preservation.} Throughout \textsc{KVonly} and \textsc{KVandW}, every increment of eviction adds equal restoration work to both paths, so the placement remains balanced wherever eviction stops. \textsc{Wonly} and \textsc{CompKV} extend beyond the point where this balance can be maintained: no KV chunks are left to assign to the decompression path, so any further eviction can no longer be balanced across the two paths. Because they are entered only after the earlier stages have balanced the placement across both paths, the system still preserves the best achievable balance for the given eviction amount.}

\subsection{Backward-out, Forward-in Policy}
\label{section:restore_inference_overlap}

While §\ref{section:dual_path} determines where to place evicted data across restore paths, this section addresses in what order to evict and restore weight layers and KV chunks. To determine this order, we must consider the sequential dependency structure of LLM inference: each layer ($L_i$) cannot begin until the preceding layer ($L_{i-1}$) completes, so eviction ordering directly determines whether restoration and prefill can proceed concurrently. Consider, for example, an LRU-based policy, widely used in OS page reclamation~\cite{Gorman2004-lru}, that evicts the earliest-accessed layers starting from $L_1$. Even if only a small amount of memory is reclaimed, prefill cannot start until $L_1$ is restored, leaving the GPU entirely idle. Under this policy, prefill is always blocked by restoration regardless of the eviction amount. This is especially problematic in mobile environments where the eviction amount is unpredictable.

\begin{figure}[t]
    \centering
    \includegraphics[width=\linewidth]{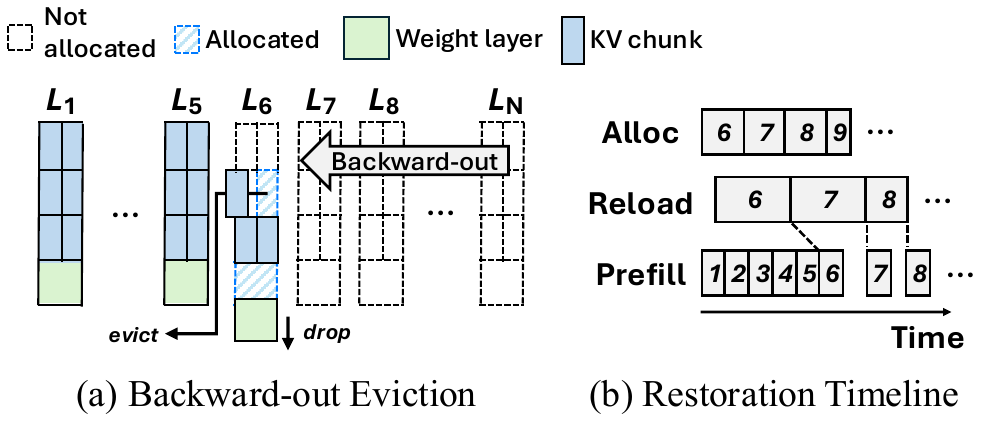}
    \caption{Backward-out eviction and restoration overlap in \ours{}. (a) Eviction proceeds from the last layer backward. (b) Timeline of GPU prefill overlapping with CPU restoration of evicted layers.}
    \label{fig:restoration_overlap}
\end{figure}

\ours{} takes the opposite approach\add{, following the well-known MRU-style ordering (Figure~\ref{fig:restoration_overlap}a)}. The system evicts \revise{from the last layer backward, starting with both weights and KV chunks of the final layer, ensuring that}{in reverse layer order, always taking the rearmost of the remaining weight layers and KV chunks.\footnote{\add{Each layer holds (context length)/256 KV chunks, each organized hidden-dimension-first.}} This \textit{backward-out} eviction ensures that} early layers are always preserved regardless of how much memory is ultimately reclaimed.\del{ We call this the backward-out eviction policy.} Restoration then proceeds forward from the earliest evicted layer (\textit{forward-in}), matching the order inference will access them. As prefill progresses through $L_1 \to L_2 \to \cdots$, the CPU concurrently restores evicted layers in the same direction, naturally overlapping the two processes and minimizing prefill stalls (Figure~\ref{fig:restoration_overlap}b).

The backward-out, forward-in policy produces a natural LIFO access pattern: compressed KV chunks are added in tail order during eviction and removed in head order during restoration.
\ours{} exploits this by organizing the in-memory swap space as a stack.
Rather than pre-allocating a large fixed-size stack, the system begins with a small stack and grows incrementally, allocating additional stack segments only when needed and releasing them after restoration completes. 
This approach minimizes the memory footprint and reduces fragmentation.
\section{\MakeUppercase{Implementation}}
\label{section:implementation}

We implement \ours{} with 6k lines of C/C++ on top of llama.cpp~\cite{Cpp2023-tx}, an open-source LLM inference framework that supports various backends including mobile GPUs~\cite{Wang2025-ct}.

\noindent\textbf{Compression and kernels.}
\ours{} treats KV cache compression as a modular component with a drop-in replaceable interface. In our implementation, we adopt two existing compression algorithms: 8-bit quantization~\cite{Sheng2023-cy} and CacheGen~\cite{Liu2024-wu}, a higher-ratio technique at the cost of slower decompression. We use 8-bit quantization as the default method, and compare the two in §\ref{section:compression_algorithm_comparison}. To enable these compression methods on the CPU, we implement new compression and decompression kernels using ARM NEON SIMD instructions~\cite{ARM2025-vm}, targeting the ARMv8-A architecture commonly found in mobile devices. Our custom GPU attention kernel is implemented in 0.6k lines of OpenCL~\cite{KhronosGroup-opencl}, and shared buffers between the CPU and GPU are realized using OpenCL's Shared Virtual Memory (SVM)~\cite{ARM2025-rg}.

\noindent\textbf{Eviction and restoration interface.}
\ours{} provides interfaces for both eviction and restoration operations, assuming that the amount of memory to release is provided externally:
\begin{itemize}[topsep=0pt, partopsep=0pt, itemsep=0pt, parsep=0pt, leftmargin=2em]
\item \texttt{evict(size):} Given the amount of memory to be released, this function evicts LLM memory based on the current state until the specified size is freed.
\item \texttt{restore\_generate(input\_tokens):} This function restores evicted LLM memory based on the current state. Using C++ primitives such as \texttt{std::promise}, it allows the GPU to begin prefill computation on layers as soon as they are restored.
\end{itemize}

% \noindent\textbf{System integration.}
% SVM buffers are allocated through the GPU driver and thus belong to 
% the page-locked region described in §\ref{section:memory_management_in_mobile_os}, 
% which the OS paging path cannot reclaim. Consequently, \ours{} operates 
% independently of OS paging path, and deploying \ours{} requires no 
% modifications to the OS kernel. For memory pressure detection, we use 
% Android's \texttt{onTrimMemory} callback in our application-level 
% deployment (§\ref{section:real_world}).
\noindent\textbf{System integration.}
\del{\ours{} operates independently of OS swap mechanisms, as SVM buffers reside in GPU memory, which is not subject to OS-level swapping.} \add{SVM buffers are allocated through the GPU driver and thus belong to the page-locked region described in §\ref{section:memory_management_in_mobile_os}, which the OS paging path cannot reclaim.} Consequently, \add{\ours{} operates independently of the OS paging path, and }deploying \ours{} requires no modifications to the OS kernel. For memory pressure detection, we use Android's \texttt{onTrimMemory} callback in our application-level deployment (§\ref{section:real_world}).

\section{\MakeUppercase{Evaluation}}

\begin{table}[t]
\centering
\caption{Evaluation testbed specifications}
\label{tab:testbed}
\resizebox{\columnwidth}{!}{
\begin{tabular}{ccccc}
\toprule
\multicolumn{5}{c}{\textbf{Devices}} \\
\midrule
\textbf{Device} & \textbf{CPU} & \textbf{GPU} & \textbf{Memory} & \textbf{Storage} \\
\midrule
Galaxy S25+ & \makecell{2×Oryon Phoenix L \\ 6×Oryon Phoenix M} & Adreno 830 & \makecell{12 GB \\ LPDDR5X} & UFS 4.0 \\
\midrule
OnePlus 12 & \makecell{1×Cortex-X4 \\ 5×Cortex-A720 \\ 2×Cortex-A520} & Adreno 750 & \makecell{12 GB \\ LPDDR5X} & UFS 4.0 \\
\toprule
\multicolumn{5}{c}{\textbf{LLM Models}} \\
\midrule
\textbf{Model} & \textbf{Weights} & \multicolumn{3}{c}{\textbf{KV cache (8k / 16k / 32k)}} \\
\midrule
Qwen3-0.6B & 1.2 GB & \multicolumn{3}{c}{0.9 GB / 1.8 GB / 3.5 GB} \\
EXAONE-4.0-1.2B & 2.6 GB & \multicolumn{3}{c}{0.5 GB / 1.0 GB / 1.9 GB} \\
\bottomrule
\end{tabular}
}
\end{table}

\begin{figure*}[!t]
\centering
\vspace{-2.3mm}
\includegraphics[width=\textwidth]{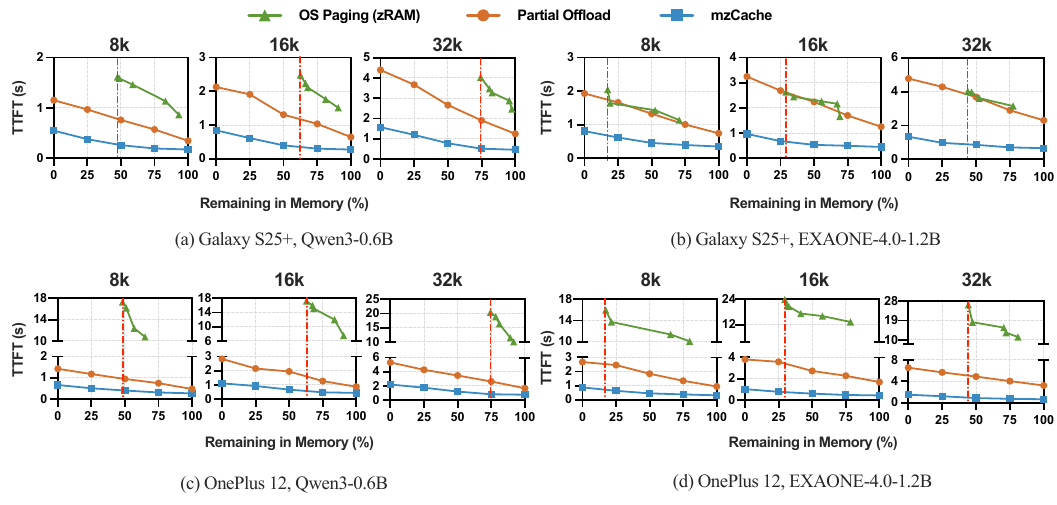}
\caption{TTFT across different remaining-memory levels. The red dashed line marks full eviction under OS Paging, which still retains non-negligible memory due to poor KV cache compression. \ours{} consistently achieves lower TTFT than baselines across all conditions.}
\label{fig:exp1}
\end{figure*}

This section evaluates \ours{}, analyzing the impact of its optimization techniques and demonstrating its practicality in mobile multitasking environments. We first compare overall performance against baselines (§\ref{section:overall_performance}), then dissect the contribution of each core technique (§\ref{section:impact_of_core_techniques}), measure power and energy consumption (§\ref{section:power}), compare compression algorithms in terms of compression ratio, accuracy, and TTFT (§\ref{section:compression_algorithm_comparison}), and validate its effectiveness in real-world deployment (§\ref{section:real_world}).

\subsection{Experimental Setup}
\label{section:exp_setup}
\noindent\textbf{Devices.} 
Table~\ref{tab:testbed} summarizes device specifications. We evaluate on two off-the-shelf smartphones representing different processor generations. On Galaxy S25+~\cite{Samsung}, big cores handle decompression and storage I/O while middle cores perform allocation. On OnePlus 12~\cite{Oneplus}, we distribute workloads across the six higher-performance cores, excluding the two little cores.

\noindent\textbf{LLM models.} 
We evaluate \ours{} using two lightweight LLMs: 
Qwen3-0.6B~\cite{Qwen-Team2025-ry} and EXAONE-4.0-1.2B~\cite{exaone-4.0}.
Table~\ref{tab:testbed} shows their weight sizes and KV cache sizes 
at different context lengths.

\noindent\textbf{Number of tokens.} 
Real-world LLM conversation traces show that accumulated 
context readily reaches thousands of tokens as interactions progress~\cite{Yan2025-sc}. We use context lengths 
of 8k, 16k, and 32k tokens to evaluate performance across 
representative KV cache sizes. We fix the input sequence 
length to 8 tokens to model a short follow-up query against 
the accumulated context.

\noindent\textbf{Baselines.} We compare \ours{} against two baselines:

\emph{1) OS Paging.} We leverage the default Android paging mechanism to represent the behavior of LLM services that rely solely on OS memory management. Model weights are managed via standard memory mapping (\texttt{mmap}) as file-backed pages, while the KV cache is treated as anonymous memory and compressed into zRAM using the zstd algorithm~\cite{facebookUnknown-wy}. Since this approach relies on OS-level page fault handling, we run inference on the CPU\revise{, with pages restored on-demand as they are accessed.}{: GPU buffers reside in the page-locked region~(\S\ref{section:memory_management_in_mobile_os}), which the OS paging path cannot reclaim, so a GPU-based deployment offers no graceful reclaim path and thus cannot serve as a partial-eviction baseline. On the CPU, pages are restored on demand as they are accessed.}\del{ This represents a best-effort approach that relies solely on OS mechanisms.}

\emph{2) Partial Offload.} Since no existing mobile LLM system provides mechanisms to handle memory pressure, we implement a partial offload baseline in llama.cpp. Note that default llama.cpp adopts monolithic memory allocation, limiting it to all-or-nothing eviction. To enable fair comparison with \ours{} across different eviction amounts, we extend it to support partial eviction and restoration at layer granularity for both weights and KV cache. Upon restoration, the baseline sequentially loads weights and KV cache from storage into GPU memory before executing prefill on the GPU.\add{ Consequently, the eviction and restoration order does not affect its performance. Only the eviction amount matters.}

\noindent\textbf{Measurements.}
We measure CPU memory usage via \texttt{smaps} and GPU memory usage through the \texttt{gpumem\_mapped\_bytes} value exposed by the Qualcomm KGSL driver~\cite{qualcomm_kgsl}. \texttt{smaps} reports how much data is swapped out, not how much space it occupies after zRAM compression. Thus, for OS Paging, we pre-profile the compression ratio from full swap-out scenarios and apply it to the reported size to estimate the actual memory footprint. Power consumption is estimated by multiplying the battery's current and voltage values obtained from \texttt{sysfs}. To evaluate the impact of different compression algorithms on accuracy, we measure F1 scores using 428 question-answer pairs from 30 documents in TriviaQA~\cite{Joshi2017-aw}, whose context lengths range from 8.8k to 28k tokens.
Except for application-level deployment experiments, all measurements are conducted through Android Debug Bridge (ADB)~\cite{adb}.

\subsection{Overall Performance - TTFT}
\label{section:overall_performance}
We evaluate restoration and inference scenarios with varying amounts 
of LLM memory remaining in system memory.
Remaining in memory denotes the percentage of weights and KV cache that stays resident in memory after eviction. This includes compressed KV cache held in system memory. We exclude token embeddings for simplicity.

For Partial Offload and \ours{}, we compare the restoration performance at 0\%, 25\%, 50\%, and 75\% of remaining memory.
Since granularity differs between the two, we allow a ±3\% margin, always evicting slightly more in \ours{} to ensure fairness.
For OS Paging, we evict data by running a memory-intensive dummy process to induce OS-level reclamation. Since the ratio of evicted weights to KV cache at a given remaining memory level can affect TTFT, we drop weights before swapping out KV cache to keep this ratio fixed across runs, isolating the remaining-memory level as the sole variable.

\noindent\textbf{TTFT comparison.}
Figure~\ref{fig:exp1} shows that \ours{} consistently achieves lower TTFT than Partial Offload across all configurations.
Specifically, \ours{} achieves a 2.1--5.5$\times$ speedup at 0\%, 25\%, 50\%, and 75\% remaining memory levels. 
The speedup is consistent regardless of KV cache size, model type, or device, demonstrating \ours{}'s effectiveness across diverse scenarios.
OS Paging performs worse than Partial Offload in most cases.
Consistent with Figure~\ref{fig:swap_slow}, this latency is dominated by on-demand paging in restoration rather than prefill computation.
On the OnePlus 12, OS Paging exhibits significantly higher latency with large variance.
During restoration, we monitored the resident set size (RSS)
and observed that it sometimes decreased rather than increased, 
indicating that on-demand page faults triggered additional evictions of already-resident pages.
This cascading effect substantially increases TTFT.

\noindent\textbf{Analysis of compressed KV cache.} 
The red vertical dashed line in Figure~\ref{fig:exp1} marks the remaining memory when all weights are dropped and all KV cache is swapped out under OS Paging. Even with full eviction, OS Paging consumes substantial memory due to poor KV cache compression. In contrast, \ours{} reduces remaining memory to 0\% by evicting compressed KV cache to storage in the \textsc{CompKV} stage. At these respective full-eviction states, mzCache achieves a 2.5--3.0$\times$ speedup over OS Paging on Galaxy S25+ and 9.2--25.9$\times$ on OnePlus 12.

\begin{figure}[!t]
    \centering
    \includegraphics[width=1\linewidth]{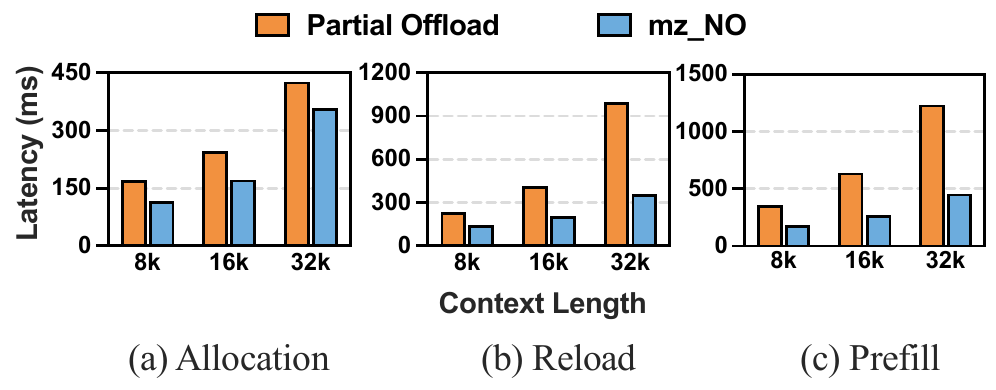}
    \caption{Latency breakdown for (a) allocation, (b) reload, and (c) prefill.
Partial Offload vs. \ours{} without overlap (mz\_NO).}
    \label{fig:exp2}
\end{figure}

\begin{figure}[!t]
\centering
% \vspace{-3mm}
\includegraphics[width=1\linewidth]{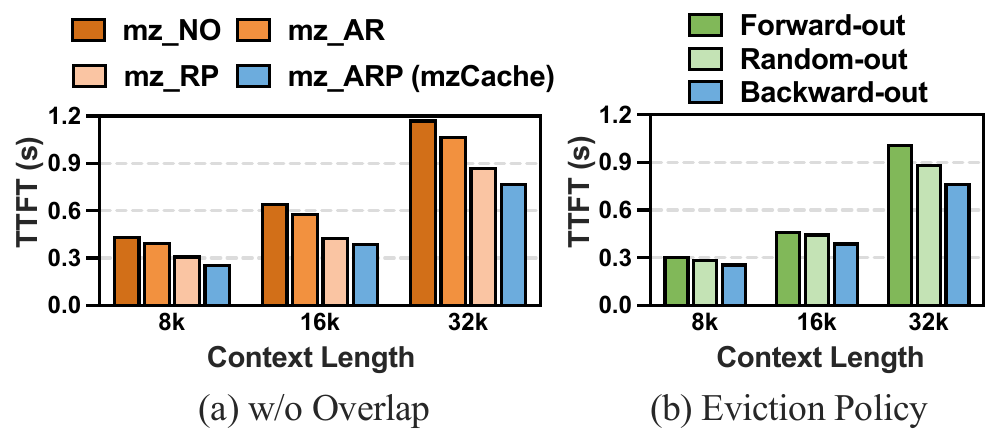}
\caption{TTFT with different (a) overlapping strategies and (b) eviction policies. mz\_NO: no overlap, mz\_AR: Allocation-Reload overlap, mz\_RP: Reload-Prefill overlap, \ours{}: Allocation-Reload-Prefill overlap. }
\label{fig:exp3}
\end{figure}

% \vspace{-0.5em}
\subsection{Impact of Core Techniques}
\label{section:impact_of_core_techniques}

To understand the contribution of each core technique in \ours{}, we conduct a detailed breakdown analysis of each technique.
For focused comparison, we use Galaxy S25+ with Qwen3-0.6B, restoring from 50\% remaining memory for a representative case unless noted.

\begin{figure}[!t]
    \centering
    \includegraphics[width=1\linewidth]{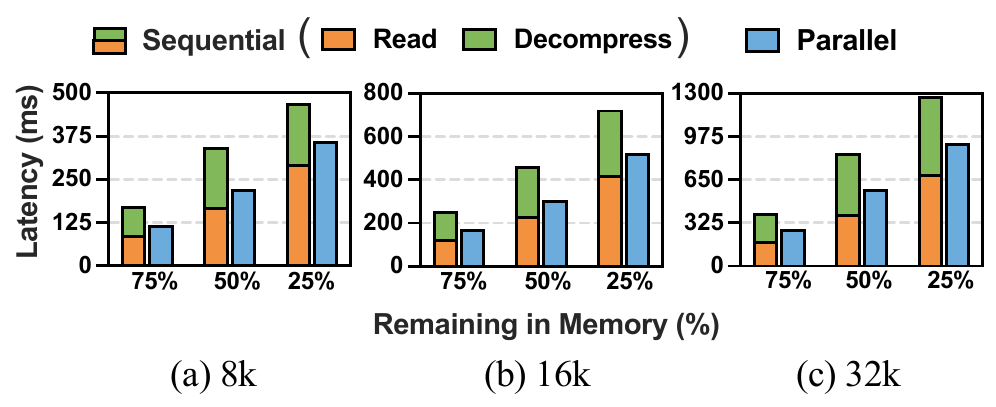}
    \caption{Sequential vs. parallel restoration time at different remaining memory levels. Sequential bars show stacked in-memory and storage restoration times.}
    \label{fig:exp_parallel}
\end{figure}

% \noindent\textbf{Individual Component Improvements.}
\noindent\textbf{Breakdown of performance improvements.}
Figure~\ref{fig:exp2} shows the latency of allocation, reload, and prefill for Partial Offload and \ours{}. 
For a clear analysis, we disable the overlapping feature of \ours{}.
\ours{} achieves a 1.37× speedup in allocation through parallelized memory allocation across CPU cores, 2.15× in reload via zero-copy transfers and hybrid swap (further analyzed in Figure~\ref{fig:exp_parallel}), and 2.33× in prefill using a mobile-optimized attention kernel.

\noindent\textbf{Benefits of overlapping.}
Figure~\ref{fig:exp3}a demonstrates the effect of overlapping by comparing four configurations: no overlap, allocation-reload overlap, reload-prefill overlap, and full overlap of all three operations. Overlapping allocation and reload achieves 1.10× speedup, while overlapping reload and prefill achieves larger gains of 1.41×. Full overlap of allocation, reload, and prefill yields 1.61× speedup, demonstrating the cumulative benefit of overlapping all phases.

\noindent\textbf{Impact of eviction ordering.}
Figure~\ref{fig:exp3}b evaluates the impact of our \emph{backward-out} eviction policy. For comprehensive comparison, we implement two alternative eviction policies: \emph{forward-out}, which evicts early layers first similar to LRU-based eviction, and \emph{random-out}, averaged over 10 runs. All policies use forward-in restoration ordering. The results show that backward-out minimizes prefill stalls during inference, confirming the importance of evicting later layers first to preserve early layers needed for immediate computation.

\noindent\textbf{Effect of hybrid swap.} 
Figure~\ref{fig:exp_parallel} shows the effect of hybrid swap, demonstrating two key benefits: first, the speedup 
achieved by restoring via parallel paths compared to sequential 
restoration, and second, the effectiveness of load balancing 
between the two restore paths. The sequential case shows that the two restoration paths achieve similar latencies, 
confirming effective load balancing. As a result, parallel 
restoration significantly outperforms the sequential case.

\subsection{Power and Energy Consumption}
\label{section:power}
\begin{figure}[!t]
\centering
% \vspace{-3mm}
\includegraphics[width=1\linewidth]{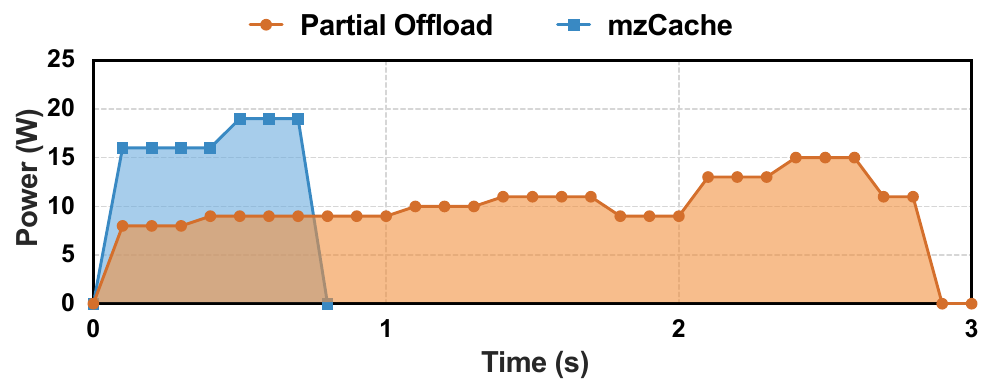}
\caption{Power consumption during restoration and prefill. The shaded area under each curve indicates total energy consumption. Due to the reduced TTFT, \ours{} consumes significantly less energy than Partial Offload despite its higher peak power.}
\label{fig:exp_power}
\end{figure}
Figure~\ref{fig:exp_power} shows the power consumption trace during restoration and prefill on Galaxy S25+ with Qwen3-0.6B at 50\% remaining memory and 32k context length. \ours{} exhibits higher peak power at 19.2 W compared to Partial Offload's 14.6 W, because \ours{} fully utilizes storage, CPU, and GPU in 
parallel. Despite the higher peak power, \ours{} completes 
the restoration and prefill process significantly faster, and as shown by the shaded area under each curve representing total energy consumption, \ours{} consumes less energy overall. This demonstrates that \ours{}'s latency reduction achieves better energy efficiency, an important consideration for battery-constrained mobile devices.

\begin{figure}[!t]
    \centering
    \includegraphics[width=1\linewidth]{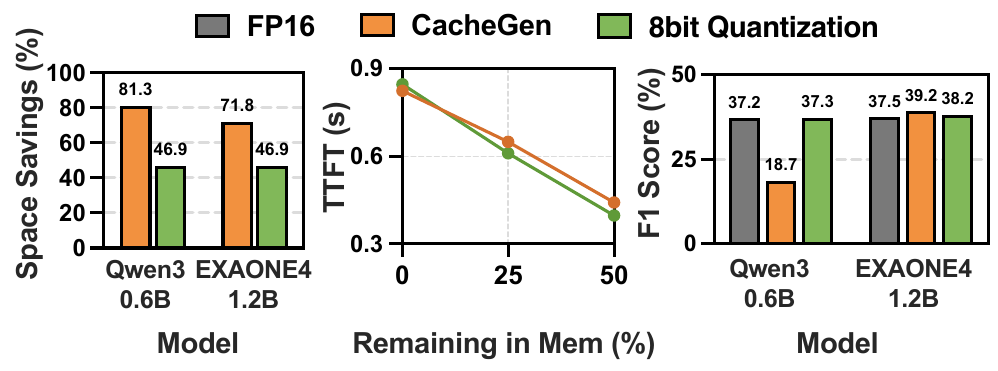}
    \caption{Comparison of 8-bit quantization and CacheGen: space savings, TTFT at different remaining memory levels, and F1 score relative to FP16.}
    \label{fig:exp_compression}
\end{figure}

\subsection{Compression Algorithm Comparison}
\label{section:compression_algorithm_comparison}

\ours{} treats KV cache compression as a modular component, 
allowing the algorithm to be replaced as needed. To analyze how the choice of compression algorithm affects performance, 
we compare 8-bit quantization~\cite{Sheng2023-cy} and 
CacheGen~\cite{Liu2024-wu} on Galaxy S25+ in Figure~\ref{fig:exp_compression}. 
Space savings and TTFT are measured at 16k context length.
At 50\%, 8-bit quantization achieves lower TTFT due to faster decompression from in-memory swap. At 0\%, CacheGen achieves better performance because its higher compression ratio reduces the size of compressed KV cache that must be loaded from storage.
Both algorithms maintain comparable F1 scores on EXAONE-4.0-1.2B, while 8-bit quantization better preserves accuracy on Qwen3-0.6B, likely due to its larger KV hidden dimension and smaller model size.
\add{Notably, both algorithms are idempotent, so repeated compression and decompression do not accumulate error, keeping accuracy stable across swap cycles.}
These results show that the choice of compression algorithm 
should be tailored to the deployment scenario.

\begin{figure}[!t]
\centering
% \vspace{-3mm}
\includegraphics[width=1\linewidth]{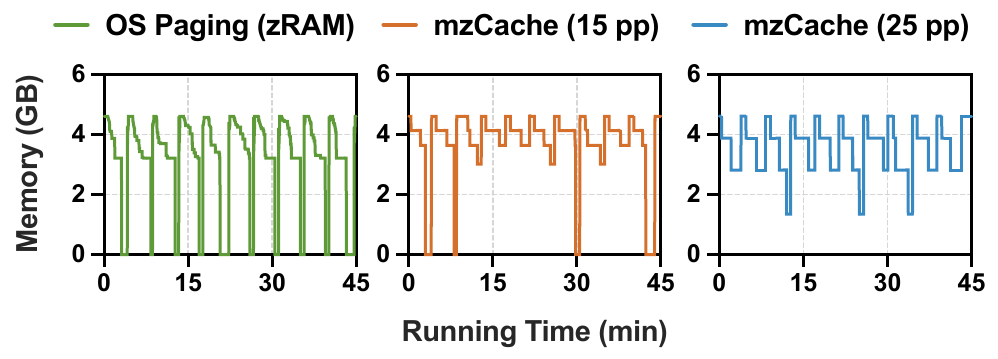}
\caption{\add{Total memory trace of an LLM application under real-world multitasking,
comparing OS Paging (zRAM) with \ours{} evicting approximately 15 pp and 25 pp
of LLM memory under pressure. OS Paging is terminated by the LMK in every round,
whereas \ours{} (25 pp) survives the entire session.}}
\label{fig:eval_real_world}
\end{figure}

\subsection{Verifying Real-world Deployment}
\label{section:real_world}
\del{To validate \ours{} in a real mobile multitasking setting, we built an Android LLM application on Galaxy S25+ using llama.cpp with GPU backend as the LLM engine. Upon receiving Android's \texttt{onTrimMemory} callback, we evict approximately 15\% of LLM memory. We created realistic pressure through typical multitasking activities, including launching the camera app and watching YouTube videos. Figure~\ref{fig:eval_real_world} shows the real-time memory traces of the LLM application with and without \ours{}. Both apps initially occupy $\sim$5 GB of memory. As multitasking begins, \ours{} detects memory pressure through its callback handler and progressively evicts memory. In contrast, the baseline app (without \ours{}), lacking any memory management capability, statically holds memory until eventually terminated by the LMK. Upon resuming the LLM application, \ours{} restores context seamlessly, whereas the baseline app requires a full cold start. This demonstrates that LLM inference systems must incorporate principled memory management to sustain operation in mobile multitasking environments. We discuss practical deployment considerations in §\ref{section:discussion}.}

\add{To validate \ours{} in a real mobile multitasking setting, we deploy an Android
LLM application on Galaxy S25+ running Qwen3-0.6B with a 32k context length.
To generate realistic memory pressure, we interleave LLM use with everyday
smartphone activity across ten rounds over $\sim$45 minutes. Within each round,
we interact with a range of popular consumer apps---social networking,
video streaming, gaming, camera, and web browsing\footnote{\add{Instagram, Facebook,
YouTube, TikTok, Netflix, Roblox, PUBG Mobile, Camera, SNOW, and Chrome.}}---before triggering a prefill on the LLM application at the round's end. We compare
three configurations: OS Paging (zRAM), from §\ref{section:exp_setup}; \ours{} (15 pp), which evicts approximately 15 percentage points (pp) of LLM
memory upon each Android \texttt{onTrimMemory} callback; and \ours{} (25 pp),
which evicts approximately 25 pp.}

\add{Figure~\ref{fig:eval_real_world} shows memory traces over the session.
OS Paging is terminated by the LMK because the KV cache compresses
poorly, forcing a cold start on every return. In contrast, \ours{} (25 pp)
progressively evicts memory as pressure builds and survives all ten rounds, restoring context on each return without a cold start. This shows that \ours{}
remains effective under real-world multitasking pressure where OS memory
management fails. With a smaller eviction amount, however, \ours{} (15 pp) is terminated in four of ten rounds. Since process termination is decided by the LMK and \ours{} operates independently of the OS, \ours{} cannot prevent such terminations entirely. How much memory to evict, and when, thus becomes important for practical deployment, which we discuss in §\ref{section:discussion}.}
\vspace{3mm}
\section{\MakeUppercase{Discussion}}
\label{section:discussion}

% \noindent\textbf{Extensibility across hardware.}
% In our implementation, the CPU handles decompression while the GPU executes prefill, but these roles are not fixed and can be adapted to different hardware configurations; for instance, prefill can run on an NPU, or decompression can be offloaded to the GPU. More broadly, although we implement \ours{} on a Qualcomm SoC, the underlying design applies to any SoC with a unified memory architecture. Porting \ours{} to another hardware platform thus requires replacing only the device-specific kernels to match the target SoC, while the rest of the system remains unchanged.
\noindent\textbf{Extensibility \revise{to heterogeneous processors}{across hardware}.}
In our implementation, the CPU handles decompression while the GPU executes prefill, but these roles are not fixed and can be adapted to different hardware configurations; for instance, prefill can run on an NPU, or decompression can be offloaded to the GPU. \revise{While such configurations may require additional implementation effort for device-specific kernels, the core design principles of \ours{} apply regardless of the processor type.}{More broadly, although we implement \ours{} on a Qualcomm SoC, its design applies to any unified-memory SoC, including those now found in laptops and workstations. Porting \ours{} to other hardware platforms thus requires replacing only the device-specific kernels to match each target SoC, while the rest of the system remains unchanged.}

% \noindent\textbf{Pressure detection and eviction amount.}
% \ours{} decouples both memory pressure detection and the eviction amount from its core design, so each can be set per deployment. An application-level deployment can detect pressure via callbacks such as \texttt{onTrimMemory} and evict a larger amount to reduce the risk of LMK termination. A system-level deployment can instead use kernel signals such as \texttt{vmpressure} or PSI, and its low \texttt{oom\_score\_adj} places it far down the LMK's kill order, substantially reducing the risk that motivated the larger eviction amount~\cite{Android2025-yo}. The core mechanism is unchanged in either case, letting \ours{} deploy across the software stack, from third-party applications to system-level services.
\noindent\textbf{\revise{Memory pressure detection}{Pressure detection and eviction amount}.}
\revise{\ours{} decouples memory-pressure detection from its core design, allowing the most appropriate mechanism to be chosen based on the deployment scenario.}{\ours{} decouples both memory pressure detection and the eviction amount from its core design, so each can be set per deployment.} \revise{For application-level deployments, OS-provided memory-pressure callbacks (e.g., \texttt{onTrimMemory}) can be used, while system-level deployments can instead rely on kernel signals such as \texttt{vmpressure} or PSI.}{An application-level deployment can detect pressure via callbacks such as \texttt{onTrimMemory} and evict a larger amount to reduce the risk of LMK termination. A system-level deployment can instead use kernel signals such as \texttt{vmpressure} or PSI, and its low \texttt{oom\_score\_adj} places it far down the LMK's kill order, substantially reducing the risk that motivated the larger eviction amount~\cite{Android2025-yo}.} \revise{This flexibility allows \ours{} to remain compatible across different levels of the software stack}{The core mechanism is unchanged in either case, making \ours{} deployable across the software stack}, from third-party applications to system-level services.

\noindent\textbf{Multi-context support.}
\ours{} currently supports a single LLM context scenario. However, as LLM applications become more prevalent, multiple LLM applications may coexist, each maintaining its own weights and KV caches. \ours{} can be extended to handle such scenarios by adding coordination policies and data structures for managing multiple weights and KV caches. Such extensions are beyond the scope of this work and we leave them for future work.

\add{
\noindent\textbf{Thermal throttling.}
\ours{} balances the two restore paths using the sequential read and decompression throughputs obtained by offline profiling. Under sustained load, thermal throttling can shift these throughputs and thereby disrupt the balance between the two paths. One way to address this is to profile across multiple memory and CPU frequency combinations in advance, allowing \ours{} to adapt to the current operating conditions at runtime. Such runtime adaptation, however, would require migrating already-evicted data between the two paths to match the new balance point.

\noindent\textbf{Selectively activated models.}
Our design assumes that first-token generation touches all model weights and the entire KV cache, which holds for dense transformer models. Some architectures, however, activate only a subset at the first token: mixture-of-experts (MoE) models route each layer to a few experts~\cite{Shazeer2017-moe}, and sparse-attention models attend to only part of the KV cache~\cite{Tang2024-quest}. For such models, the natural extension is to restore the activated subset first and defer the rest. Since which experts or KV chunks are active is input-dependent and not fully known before prefill, it must be predicted in advance, e.g., from statistics of prior interactions. Incorporating such predictions into the restoration order is an interesting direction for future work.
}
\section{\MakeUppercase{Related work}}

\noindent\textbf{Offloading-based LLM inference.} Several systems offload LLM data between GPU, CPU, or disk \revise{during active}{throughout} inference to accommodate model weights and KV cache that exceed GPU memory~\cite{Sheng2023-cy, Lee2024-infinigen, Song2024-powerinfer}.\del{ These systems distribute weights, KV cache, or selectively activated neurons across memory tiers, scheduling data movement based on profiled hardware characteristics and known memory capacities.} Other work focuses on managing stored KV cache contexts across memory tiers to accelerate restoration upon reuse~\cite{Yu2025-kg, Gao2025-ey, Chen2025-ub, Jeong2025-ll, Yin2024-uo}. In all cases, the LLM system controls what to offload and when, with full knowledge of available memory. \ours{} addresses a fundamentally different setting where eviction is externally imposed under unpredictable memory pressure, requiring \revise{dynamic strategies that adapt to unknown eviction amounts rather than offloading within a fixed memory budget}{a new system that adapts its memory management to unknown eviction amounts at runtime}.

\noindent\textbf{Mobile LLM inference systems.} Existing optimizations for on-device LLM inference focus on accelerating runtime execution through heterogeneous co-execution across mobile processors~\cite{Chen2025-fe, Xu2025-ji, Hao2025-npu}, compiler and system stack optimizations for mobile SoCs~\cite{mlc-llm, proc:osdi22:walle}, and storage-based weight offloading for models exceeding device memory~\cite{Xue2024-powerinfer2, Wang2025-neuralink, Yi2025-um}. \revise{These systems target active runtime inference.}{However, these are all runtime inference optimizations.} \revise{However, mobile devices are inherently multitasking environments, and}{In mobile multitasking environments,} preserving the KV cache across interactions to \revise{maintain conversational context introduces}{reuse prior context creates} a memory management challenge \revise{that extends beyond inference into idle periods}{in the idle periods between inferences, a phase these systems do not manage}. \ours{} is the first system to address \revise{this challenge}{it}.

\noindent\textbf{Hybrid swap in mobile OS.} In the mobile OS domain, some systems combine in-memory compressed swap with storage-backed swap \del{in a tiered manner }to sustain responsiveness under memory pressure~\cite{Lim2023-dg, Liang2025-te, son2021, lebeck2020}\add{, an idea that inspired our design}. However, \del{we show that }their general-purpose compression fails on \add{the }LLM KV cache (§\ref{section:memory_management_in_mobile_os}). \revise{\ours{} adapts the hybrid swap concept specifically for LLM inference,}{Rather than modifying the OS, we adapt hybrid swap within the LLM inference system itself, specializing it for the KV cache and} \revise{utilizing}{driving} both paths in parallel with balanced load distribution\del{ and requiring no}\add{, without} OS kernel modifications.
\vspace{3mm}
\section{\MakeUppercase{Conclusion}}
In this paper, we present \ours{}, an on-device LLM inference system that elastically evicts and rapidly restores LLM memory under external memory pressure. Our evaluation on commercial smartphones shows that \ours{} achieves 2.1--5.5$\times$ faster TTFT compared to storage-backed partial offload, and preserves the user's context under the pressure of real-world app usage. These results validate that dedicated memory management, rather than reliance on OS-based mechanisms, is essential for responsive on-device LLM inference in multitasking environments.

\vspace{3.5mm}
\section*{ACKNOWLEDGMENTS}
We sincerely thank the anonymous reviewers and our shepherd for their
valuable comments that helped improve this paper.
This work was supported by the National Research Foundation of Korea (NRF) grant (RS-2022-NR070834), an Institute of Information \& Communications Technology Planning \& Evaluation (IITP)-Information Technology Research Center (ITRC) grant (IITP-2021-0-02048), and an IITP grant (RS-2024-00418784), funded by the Korea government (MSIT).
Kyunghan Lee is the corresponding author.

\newpage

\bibliographystyle{ACM-Reference-Format}
\bibliography{reference}

@inproceedings{son2021,
  title={ASAP: Fast Mobile Application Switch via Adaptive Prepaging},
    author = {Son, Sam and Lee, Seung Yul and Bae, Jonghyun and Jin, Yunho and Jeong, Jinkyu and Ham, Tae Jun and Lee, Jae W. and Yoon, Hongil},
  booktitle={Proceedings of the 2021 USENIX Annual Technical Conference},
  pages={365--380},
  year={2021}
}

@inproceedings{proc:osdi22:walle,
  title={Walle: An End-to-End, General-Purpose, and Large-Scale Production System for Device-Cloud Collaborative Machine Learning},
  author={Chengfei Lv and Chaoyue Niu and Renjie Gu and Xiaotang Jiang and Zhaode Wang and Bin Liu and Ziqi Wu and Qiulin Yao and Congyu Huang and Panos Huang and Tao Huang and Hui Shu and Jinde Song and Bin Zou and Peng Lan and Guohuan Xu and Fei Wu and Shaojie Tang and Fan Wu and Guihai Chen},
  booktitle={Proceedings of the 16th USENIX Symposium on Operating Systems Design and Implementation},
  pages={249--265},
  year={2022}
}

@inproceedings{lebeck2020,
  title={End the Senseless Killing: Improving Memory Management for Mobile Operating Systems},
  author={Niel Lebeck and Arvind Krishnamurthy and Henry M. Levy and Irene Zhang},
  booktitle={Proceedings of the 2020 USENIX Annual Technical Conference},
  pages={873--887},
  year={2020}
}

@inproceedings{Chen2025-ub,
  title={IMPRESS: An importance-informed multi-tier prefix KV storage system for large language model inference},
  author={Chen, Weijian and He, Shuibing and Qu, Haoyang and Zhang, Ruidong and Yang, Siling and Chen, Ping and Zheng, Yi and Huai, Baoxing and Chen, Gang},
  booktitle={Proceedings of the 23rd USENIX Conference on File and Storage Technologies},
  pages={187--201},
  year={2025}
}

@inproceedings{Kwon2023-ci,
  title={Efficient memory management for large language model serving with PagedAttention},
  author={Kwon, Woosuk and Li, Zhuohan and Zhuang, Siyuan and Sheng, Ying and Zheng, Lianmin and Yu, Cody Hao and Gonzalez, Joseph and Zhang, Hao and Stoica, Ion},
  booktitle={Proceedings of the ACM SIGOPS 29th Symposium on Operating Systems Principles},
  pages={611--626},
  year={2023}
}

@inproceedings{Jeong2025-ll,
  title={Accelerating LLM serving for multi-turn dialogues with efficient resource management},
  author={Jeong, Jinwoo and Ahn, Jeongseob},
  booktitle={Proceedings of the 30th ACM International Conference on Architectural Support for Programming Languages and Operating Systems},
  pages={1--15},
  year={2025}
}

@inproceedings{Wang2024-or,
  title={MNN-LLM: A generic inference engine for fast large language model deployment on mobile devices},
  author={Wang, Zhaode and Yang, Jingbang and Qian, Xinyu and Xing, Shiwen and Jiang, Xiaotang and Lv, Chengfei and Zhang, Shengyu},
  booktitle={Proceedings of the 6th ACM International Conference on Multimedia in Asia Workshops},
  pages={1--7},
  year={2024}
}

@inproceedings{Gao2025-ey,
  title={Fast state restoration in LLM serving with HCache},
  author={Gao, Shiwei and Chen, Youmin and Shu, Jiwu},
  booktitle={Proceedings of the 20th European Conference on Computer Systems},
  pages={128--143},
  year={2025}
}

@inproceedings{Sheng2023-cy,
  title={FlexGen: High-throughput generative inference of large language models with a single GPU},
  author={Sheng, Ying and Zheng, Lianmin and Yuan, Binhang and Li, Zhuohan and Ryabinin, Max and Chen, Beidi and Liang, Percy and Ré, Christopher and Stoica, Ion and Zhang, Ce},
  booktitle={Proceedings of the 40th International Conference on Machine Learning},
  pages={31094--31116},
  year={2023}
}

@inproceedings{Lim2023-dg,
  title={SWAM: Revisiting swap and OOMK for improving application responsiveness on mobile devices},
  author={Lim, Geunsik and Kang, Donghyun and Ham, Myungjoo and Eom, Young Ik},
  booktitle={Proceedings of the 29th Annual International Conference on Mobile Computing and Networking},
  pages={1--15},
  year={2023}
}

@inproceedings{Xu2025-ji,
  title={Fast on-device LLM inference with NPUs},
  author={Xu, Daliang and Zhang, Hao and Yang, Liming and Liu, Ruiqi and Huang, Gang and Xu, Mengwei and Liu, Xuanzhe},
  booktitle={Proceedings of the 30th ACM International Conference on Architectural Support for Programming Languages and Operating Systems},
  pages={445--462},
  year={2025}
}

@inproceedings{Liu2024-wu,
  title={CacheGen: KV cache compression and streaming for fast large language model serving},
  author={Liu, Yuhan and Li, Hanchen and Cheng, Yihua and Ray, Siddhant and Huang, Yuyang and Zhang, Qizheng and Du, Kuntai and Yao, Jiayi and Lu, Shan and Ananthanarayanan, Ganesh and Maire, Michael and Hoffmann, Henry and Holtzman, Ari and Jiang, Junchen},
  booktitle={Proceedings of the ACM SIGCOMM 2024 Conference},
  pages={38--56},
  year={2024}
}

@inproceedings{Yu2025-kg,
  title={Stateful large language model serving with pensieve},
  author={Yu, Lingfan and Lin, Jinkun and Li, Jinyang},
  booktitle={Proceedings of the 20th European Conference on Computer Systems},
  pages={144--158},
  year={2025}
}

@inproceedings{Joshi2017-aw,
  title={TriviaQA: A large scale distantly supervised challenge dataset for reading comprehension},
  author={Joshi, Mandar and Choi, Eunsol and Weld, Daniel and Zettlemoyer, Luke},
  booktitle = {Proceedings of the 55th Annual Meeting of the Association for Computational Linguistics},
  pages = {1601--1611},
  year={2017}
}

@inproceedings{Chen2025-fe,
  title={Characterizing mobile SoC for accelerating heterogeneous LLM inference},
  author={Chen, Le and Feng, Dahu and Feng, Erhu and Wang, Yingrui and Zhao, Rong and Xia, Yubin and Xu, Pinjie and Chen, Haibo},
  booktitle={Proceedings of the ACM SIGOPS 31st Symposium on Operating Systems Principles},
  pages={359--374},
  year={2025}
}

@inproceedings{Hooper2024-yl,
  title={KVQuant: Towards 10 Million Context Length LLM Inference with KV Cache Quantization},
    author={Hooper, Coleman and Kim, Sehoon and Mohammadzadeh, Hiva and Mahoney, Michael W. and Shao, Yakun Sophia and Keutzer, Kurt and Gholami, Amir},
  booktitle={Proceedings of the Neural Information Processing Systems Conference},
  year={2024}
}

@inproceedings{li2025mobilora,
  title={MobiLoRA: Accelerating LoRA-based LLM Inference on Mobile Devices via Context-aware KV Cache Optimization},
  author={Li, Borui and Wang, Yitao and Ma, Haoran and Chen, Ligeng and Xiao, Jun and Wang, Shuai},
  booktitle={Proceedings of the 63rd Annual Meeting of the Association for Computational Linguistics},
  pages={23400--23410},
  year={2025}
}

@inproceedings{Song2024-powerinfer,
  title={PowerInfer: Fast Large Language Model Serving with a Consumer-grade GPU},
  author={Song, Yixin and Mi, Zeyu and Xie, Haotong and Chen, Haibo},
  booktitle={Proceedings of the ACM SIGOPS 30th Symposium on Operating Systems Principles},
  pages={590--606},
  year={2024}
}

@inproceedings{Lee2024-infinigen,
  title={InfiniGen: Efficient Generative Inference of Large Language Models with Dynamic KV Cache Management},
  author={Lee, Wonbeom and Lee, Jungi and Seo, Junghwan and Sim, Jaewoong},
  booktitle={Proceedings of the 18th USENIX Symposium on Operating Systems Design and Implementation},
  pages={155--172},
  year={2024}
}

@inproceedings{Wang2025-neuralink,
  title={Neuralink: Fast on-Device LLM Inference with Neuron Co-Activation Linking},
  author={Wang, Tuowei and Fan, Ruwen and Huang, Minxing and Hao, Zixu and Li, Kun and Cao, Ting and Lu, Youyou and Zhang, Yaoxue and Ren, Ju},
  booktitle={Proceedings of the 30th ACM International Conference on Architectural Support for Programming Languages and Operating Systems},
  pages={147--162},
  year={2025}
}

@inproceedings{Liu2024-kivi,
  title={{KIVI}: A Tuning-Free Asymmetric 2bit Quantization for {KV} Cache},
  author={Liu, Zirui and Yuan, Jiayi and Jin, Hongye and Zhong, Shaochen and Xu, Zhaozhuo and Braverman, Vladimir and Chen, Beidi and Hu, Xia},
  booktitle={Proceedings of the 41st International Conference on Machine Learning},
  pages={32332--32344},
  year={2024}
}

@inproceedings{Zhang2024-cq,
  title={{KV} Cache is 1 Bit Per Channel: Efficient Large Language Model Inference with Coupled Quantization},
  author={Zhang, Tianyi and Yi, Jonah and Xu, Zhaozhuo and Shrivastava, Anshumali},
  booktitle={Proceedings of the Neural Information Processing Systems Conference},
  year={2024}
}

@inproceedings{Zheng2024-sglang,
  title={SGLang: Efficient Execution of Structured Language Model Programs},
  author={Zheng, Lianmin and Yin, Liangsheng and Xie, Zhiqiang and Sun, Chuyue and Huang, Jeff and Yu, Cody Hao and Cao, Shiyi and Kozyrakis, Christos and Stoica, Ion and Gonzalez, Joseph E. and Barrett, Clark and Sheng, Ying},
  booktitle={Proceedings of the Neural Information Processing Systems Conference},
  year={2024}
}

@inproceedings{vaswani2017attention,
  title={Attention is all you need},
  author={Vaswani, Ashish and Shazeer, Noam and Parmar, Niki and Uszkoreit, Jakob and Jones, Llion and Gomez, Aidan N. and Kaiser, {\L}ukasz and Polosukhin, Illia},
  booktitle={Proceedings of the Neural Information Processing Systems Conference},
  year={2017}
}

@inproceedings{Dao2022-pj,
  title   = {FlashAttention: Fast and memory-efficient exact attention with IO-awareness},
  author  = {Dao, Tri and Fu, Daniel Y and Ermon, Stefano and Rudra, Atri and R\'{e}, Christopher},
  booktitle = {Proceedings of the Neural Information Processing Systems Conference},
  year    = {2022}
}

@inproceedings{Tang2024-quest,
  title={{QUEST}: Query-Aware Sparsity for Efficient Long-Context {LLM} Inference},
  author={Tang, Jiaming and Zhao, Yilong and Zhu, Kan and Xiao, Guangxuan and Kasikci, Baris and Han, Song},
  booktitle={Proceedings of the 41st International Conference on Machine Learning},
  pages={47901--47911},
  year={2024}
}

@inproceedings{Shazeer2017-moe,
  title={Outrageously Large Neural Networks: The Sparsely-Gated Mixture-of-Experts Layer},
  author={Shazeer, Noam and Mirhoseini, Azalia and Maziarz, Krzysztof and Davis, Andy and Le, Quoc and Hinton, Geoffrey and Dean, Jeff},
  booktitle={Proceedings of the 5th International Conference on Learning Representations},
  year={2017}
}

@misc{Samsung,
  author={Samsung},
  title={Samsung Galaxy S25+},
  url={https://www.samsung.com/us/smartphones/galaxy-s25/},
  year={2025}
}

@misc{Oneplus,
  author={OnePlus},
  title={OnePlus 12},
  url={https://www.oneplus.com/global/12},
  year={2024}
}

@misc{Oneplus15,
  author={OnePlus},
  title={OnePlus 15},
  url={https://www.oneplus.com/global/15},
  year={2025}
}

@misc{iPhone17,
  author={Apple},
  title={iPhone 17},
  url={https://www.apple.com/iphone-17/},
  year={2025}

}

@misc{yi2023mllm,
  title={mllm: fast and lightweight multimodal LLM inference engine for mobile and edge devices},
  author={Rongjie Yi and Xiang Li and Zhenyan Lu and Hao Zhang and Daliang Xu and Liming Yang and Weikai Xie and Chenghua Wang and Xuanzhe Liu and Mengwei Xu},
  year={2023},
  url={https://github.com/UbiquitousLearning/mllm}
}

@software{mlc-llm,
  author={MLC Team},
  title={MLC-LLM},
  url={https://github.com/mlc-ai/mlc-llm},
  year={2023}
}

@misc{Triggs2024-cf,
  title={Tested: Google's Pixel 9 Pro models only have 13GB RAM for apps},
  author={Triggs, Robert},
  booktitle={Android Authority},
  year={2024},
  url={https://www.androidauthority.com/tested-pixel-9-pro-ai-ram-3472624/}
}

@misc{Android2023-ma,
  title={Overview of memory management},
  author={Android},
  booktitle={Android Developers},
  year={2023},
  url={https://developer.android.com/topic/performance/memory-overview}
}

@misc{adb,
  title={Android Debug Bridge},
  author={Google},
  booktitle={Android Debug Bridge},
  url={https://developer.android.com/studio/command-line/adb},
  year={2025}
}

@misc{Android2025-yo,
  title={Memory allocation among processes},
  author={Android},
  booktitle={Android Developers},
  year={2025},
  url={https://developer.android.com/topic/performance/memory-management}
}

@misc{Apple2025-eu,
  title={Updates to Apple's on-device and server foundation language models},
  author={Apple},
  booktitle={Apple Machine Learning Research},
  year={2025},
  url={https://machinelearning.apple.com/research/apple-foundation-models-2025-updates}
}

@misc{ARM2025-hw,
  title={Initializing the SMMU},
  author={ARM},
  year={2025},
  url={https://developer.arm.com/documentation/100310/0202/Software-initialization-examples/Initializing-the-SMMU}
}

@misc{Cpp2023-tx,
  title={llama.cpp: LLM inference in C/C++},
  author={Gerganov, Georgi},
  year={2023},
  url={https://github.com/ggml-org/llama.cpp}
}

@misc{Huawei2025-ri,
  title={HiAI - HiAI IDE},
  author={Huawei},
  booktitle={Huawei Developer},
  year={2025},
  url={https://developer.huawei.com/consumer/en/hiai#Engine}
}

@misc{Android2025-vb,
  title={Gemini nano},
  author={Android},
  booktitle={Android Developers},
  year={2025},
  url={https://developer.android.com/ai/gemini-nano}
}

@misc{Samsung2025-tp,
  title={Galaxy AI},
  author={Samsung},
  booktitle={Samsung US},
  year={2025},
  url={https://www.samsung.com/us/galaxy-ai/}
}

@misc{ARM2025-rg,
  title={Shared virtual memory},
  author={ARM},
  booktitle={ARM Developer},
  year={2025},
  url={https://developer.arm.com/documentation/101574/0601/OpenCL-2-0/Shared-virtual-memory}
}

@misc{Android2025-xo,
  title={perfetto},
  author={Android},
  booktitle={Android Developers},
  year={2025},
  url={https://developer.android.com/tools/perfetto}
}

@misc{ARM2025-vm,
  title={Neon Intrinsics},
  author={ARM},
  year={2025},
  url={https://developer.arm.com/architectures/instruction-sets/intrinsics/}
}

@misc{Triggs2025-le,
  title={The Pixel 10 comes with 12GB of RAM, but Google has locked some of it away},
  author={Triggs, Robert},
  booktitle={Android Authority},
  year={2025},
  url={https://www.androidauthority.com/pixel-10-ai-ram-use-3591327/}
}

@misc{Wang2025-ct,
  title={Introducing the new OpenCL GPU backend in llama.cpp for Qualcomm Adreno GPUs},
  author={Wang, Hongqiang},
  booktitle={Qualcomm Developer Blog},
  year={2024},
  url={https://www.qualcomm.com/developer/blog/2024/11/introducing-new-opn-cl-gpu-backend-llama-cpp-for-qualcomm-adreno-gpu}
}

@misc{Hawkes2020-zf,
  title={Attacking the Qualcomm Adreno GPU},
  author={Hawkes, Ben},
  booktitle={Project Zero},
  year={2020},
  url={https://googleprojectzero.blogspot.com/2020/09/attacking-qualcomm-adreno-gpu.html}
}

@misc{lz4Unknown-kw,
  title={LZ4: Extremely Fast Compression Algorithm},
  author={Collet, Yann},
  url={https://github.com/lz4/lz4},
  year={2011}
}

@misc{facebookUnknown-wy,
  author={Collet, Yann},
  title={zstd: Zstandard - Fast real-time compression algorithm},
  url={https://github.com/facebook/zstd},
  year={2016}
}

@misc{ShareGPT2022-wf,
  author={ShareGPT},
  title={ShareGPT},
  year={2022},
  url={https://sharegpt.com/}
}

@online{qualcomm_kgsl,
  author = {Qualcomm},
  title = {Qualcomm Linux Graphics Guide},
  year = {2024},
  url = {https://docs.qualcomm.com/doc/80-70014-19/topic/graphics-overview.html}
}

@misc{Gorman2004-lru,
  title={Page Frame Reclamation},
  author={Gorman, Mel},
  url={https://www.kernel.org/doc/gorman/html/understand/understand013.html},
  year={2004}
}

@misc{Lam2024-adreno,
  author = {Chester Lam},
  title  = {The Snapdragon X Elite's Adreno iGPU},
  year   = {2024},
  url    = {https://chipsandcheese.com/p/the-snapdragon-x-elites-adreno-igpu}
}

@misc{KhronosGroup-opencl,
  author = {{Khronos Group}},
  title  = {OpenCL -- The Open Standard for Parallel Programming of Heterogeneous Systems},
  year   = {2025},
  url    = {https://www.khronos.org/opencl/}
}

@inproceedings{Qwen-Team2025-ry,
  title   = {Qwen3 Technical Report},
  author  = {{Qwen Team}},
  booktitle = {arXiv preprint arXiv:2505.09388},
  year    = {2025}
}

@inproceedings{Liang2025-te,
  title={Ariadne: A Hotness-Aware and Size-Adaptive Compressed Swap Technique for Fast Application Relaunch and Reduced CPU Usage on Mobile Devices},
  author={Liang, Yu and Shen, Aofeng and Xue, Chun Jason and Pan, Riwei and Mao, Haiyu and Ghiasi, Nika Mansouri and Jiang, Qingcai and Nadig, Rakesh and Li, Lei and Ausavarungnirun, Rachata and Sadrosadati, Mohammad and Mutlu, Onur},
  booktitle={Proceedings of the 2025 IEEE International Symposium on High-Performance Computer Architecture},
  pages={1588--1602},
  year={2025}
}

@inproceedings{Yin2024-uo,
  title   = {LLM as a System Service on Mobile Devices},
  author  = {Yin, Wangsong and Xu, Mengwei and Li, Yuanchun and Liu, Xuanzhe},
  booktitle = {arXiv preprint arXiv:2403.11805},
  year    = {2024}
}

@inproceedings{Gemma-Team2024-lv,
  title   = {Gemma 2: Improving open language models at a practical size},
  author  = {Gemma Team},
  booktitle = {arXiv preprint arXiv:2408.00118},
  year    = {2024}
}

@inproceedings{Zheng2023-cn,
  title   = {LMSYS-Chat-1M: A Large-Scale Real-World LLM Conversation Dataset},
  author  = {Zheng, Lianmin and Chiang, Wei-Lin and Sheng, Ying and Li,
                   Tianle and Zhuang, Siyuan and Wu, Zhanghao and Zhuang,
                   Yonghao and Li, Zhuohan and Lin, Zi and Xing, Eric P. and
                   Zhang, Hao and Gonzalez, Joseph E. and Stoica, Ion},
  booktitle = {arXiv preprint arXiv:2309.11998},
  year    = {2023}
}

@inproceedings{exaone-4.0,
  title={EXAONE 4.0: Unified Large Language Models Integrating Non-reasoning and Reasoning Modes},
  author={{LG AI Research}},
  booktitle={arXiv preprint arXiv:2507.11407},
  year={2025}
}

@inproceedings{dubey2024llama,
  title={The Llama 3 herd of models},
  author={{Llama Team}},
  booktitle={arXiv preprint arXiv:2407.21783},
  year={2024}
}

@inproceedings{zheng2024edge_llm,
  title={A Review on Edge Large Language Models: Design, Execution, and Applications},
  author={Zheng, Yue and Chen, Yuhao and Qian, Bin and Shi, Xiufang and Shu, Yuanchao and Chen, Jiming},
  booktitle={arXiv preprint arXiv:2410.11845},
  year={2024}
}

@inproceedings{Yan2025-sc,
  title     = {ShareChat: A Dataset of Chatbot Conversations in the Wild},
  author    = {Yan, Yueru and Nguyen, Tuc and Su, Bo and Lieffers, Melissa and Le, Thai},
  booktitle = {arXiv preprint arXiv:2512.17843},
  year      = {2025}
}

@inproceedings{Hao2025-npu,
  title={Scaling LLM Test-Time Compute with Mobile NPU on Smartphones},
  author={Hao, Zixu and Wei, Jianyu and Wang, Tuowei and Huang, Minxing and Jiang, Huiqiang and Jiang, Shiqi and Cao, Ting and Ren, Ju},
  booktitle={arXiv preprint arXiv:2509.23324},
  year={2025}
}

@inproceedings{Xue2024-powerinfer2,
  title={PowerInfer-2: Fast Large Language Model Inference on a Smartphone},
  author={Xue, Zhenliang and Song, Yixin and Mi, Zeyu and Zheng, Xinrui and Xia, Yubin and Chen, Haibo},
  booktitle={arXiv preprint arXiv:2406.06282},
  year={2024}
}

@article{Chen2025-pg,
  title={LLM for mobile: An initial roadmap},
  author={Chen, Daihang and Liu, Yonghui and Zhou, Mingyi and Zhao, Yanjie and Wang, Haoyu and Wang, Shuai and Chen, Xiao and Bissyandé, Tegawendé F. and Klein, Jacques and Li, Li},
  journal={ACM Transactions on Software Engineering and Methodology},
  year={2025}
}

@article{Deng2019-oz,
  title={Measuring smartphone usage and task switching with log tracking and self-reports},
  author={Deng, Tao and Kanthawala, Shaheen and Meng, Jingbo and Peng, Wei and Kononova, Anastasia and Hao, Qi and Zhang, Qinhao and David, Prabu},
  journal={Mobile Media \& Communication},
  year={2019}
}

@article{Yi2025-um,
  title={EdgeMoE: Empowering sparse large language models on mobile devices},
  author={Yi, Rongjie and Guo, Liwei and Wei, Shiyun and Zhou, Ao and Wang, Shangguang and Xu, Mengwei},
  journal={IEEE Transactions on Mobile Computing},
  year={2025}
}

@book{silberschatz2018operating,
  title={Operating System Concepts},
  author={Silberschatz, Abraham and Galvin, Peter B. and Gagne, Greg},
  year={2018},
  publisher={Wiley},
  edition={10th}
}

\end{document}